\documentclass[10pt,a4paper,aps,pra,superscriptaddress,onecolumn,floatfix]{revtex4}

\usepackage{fullpage}
\usepackage{textcomp}
\usepackage{tabularx}
\usepackage{booktabs}
\usepackage{array}
\usepackage{bbm}
\usepackage{bm}
\usepackage{xcolor}

\usepackage{tikz}

\usepackage[a4paper]{geometry}
\usepackage[utf8]{inputenc}
\usepackage{graphicx,amsmath,amssymb,amsthm} 
\usepackage[charter,cal=cmcal,sfscaled=false]{mathdesign}
\usepackage{hyperref}

\begin{document}

\title{Self-coherent phase reference sharing for continuous-variable quantum key distribution}

\author{Adrien~Marie}
\email{adrien.marie@telecom-paristech.fr}
\affiliation{T\'el\'ecom ParisTech, LTCI, CNRS, 46 Rue Barrault, 75013 Paris, France}

\author{Romain~All\'eaume}
\affiliation{T\'el\'ecom ParisTech, LTCI, CNRS, 46 Rue Barrault, 75013 Paris, France}
\date{\today}

\begin{abstract} 
We develop a comprehensive framework to model and optimize the performance of CV-QKD with a local local oscillator (LLO), when phase reference sharing and QKD are conjointly implemented with the same hardware.
We first analyze the limitations of the only existing approach, called LLO-sequential, and show that it requires high modulation dynamics and can only tolerate small phase noise, leading to expensive hardware requirements. 
Our main contribution is to introduce two original designs  to perform LLO CV-QKD  with shared hardware,, respectively called LLO-delayline and LLO-displacement, and to study their performance. Both designs rely on a self-coherent approach, in which phase reference information and quantum information are coherently obtained from a single optical wavefront. 
 
We show that these designs can lift some important limitations of the existing LLO-sequential approach. The LLO-delayline design can in particular tolerate much stronger phase noise and thus appears as an appealing alternative to LLO-sequential that can moreover be deployed with affordable hardware. We also investigate, with the LLO-displacement design, how phase reference information and quantum information can be multiplexed in a single optical pulse. By studying the trade-off between phase reference recovery and phase noise induced by displacement we however demonstrate that this design can only tolerate low phase noise. On the other hand, the LLO-displacement design has the advantage of minimal hardware requirements and can be applied to multiplex classical and quantum communications, opening practical path towards the development of coherent quantum communications systems compatible with next-generation networks requirements.
\end{abstract}

\maketitle 

\section{Introduction}

Quantum key distribution (QKD) \cite{QKDSecurity1,QKDSecurity2,DiamantiLoQi2016} is a promising technology that has reached the commercialization step since the last decade \cite{sequrenet, idquantique}. Targeting deployment over large-scale networks, next-generation QKD should rely on affordable optical components. It will in particular consist in highly integrated systems able to operate at high rate and to be deployed over modern optical networks. Relying on standard telecommunication equipment, Continuous-Variable (CV) QKD is an attractive approach towards this new step of QKD development \cite{GaussianQuantumInformation, ReviewCVQKD}. While first results towards CV-QKD practical photonics chip integration have been pursued \cite{PhotonicsIntegration,OrieuxDiamanti2016}, the possibility to effectively deploy CV-QKD in coexistence with intense wavelength-division multiplexing classical channels has been demonstrated \cite{wdm}. Furthermore, high repetition rates (up to the order of hundreds of MHz) \cite{YuemengChi, ExpCVQKD-chinois} CV-QKD systemps have also been demonstrated recently. More sensitive to optical losses than discrete-variable based QKD, long distance CV-QKD has however been demonstrated by controlling excess noise \cite{LongDistanceCVQKD1} and developping high efficiency error correction codes \cite{LongDistanceCVQKD2}.  
These important steps in the recent development of CV-QKD are, in 	addition, likely to benefit from the rise of classical coherent communications \cite{kikuchi2016fundamentals}, with the prospect of d convergence of classical and quantum communication techniques and simplified photonic integration. This positions CV-QKD and more generally quantum coherent communications as an appealing technology for the the development of modern quantum communications.

 If we compare with discrete-variable quantum communications, quantum coherent communications however have to address one specific challenge, namely phase reference sharing.
As a matter of fact the receiver must perform a phase-sensitive detection, using an optical beam usually called ``local oscillator'' whose phase drift with respect to the emitter must be controlled, or estimated and corrected. We will review in Sec. \ref{Sec: Previous Work} the different methods that have been considered to perform phase reference sharing in CV-QKD, and explain why generating "locally" the local oscillator is a fundamental requirement for continuous-variable quantum key distribution (CV-QKD), both for performance and security reasons. We will first analyze the question of phase reference sharing within to the broader body of work on reference frame agreement and then focus more specifically on the issue of phase reference sharing in coherent optical communications. \\



\textbf{Sharing a reference frame.} A reference frame, shared or partially shared, between an emitter and a receiver is a typical requirement in communication protocols, even though this requirement is often implicit. Information on the reference frame allows the receiver to more faithfully translate the received physical signals into logical information. It can for example consist in the knowledge of the relative angle between spatial two-dimensional cartesian reference frames \cite{PeresScudo2002}, in the synchronization of spatially separated clocks \cite{einstein1905elektrodynamik}, or information about the relative phase between two lasers, respectively at emitter and receiver side, when coherent optical communication is performed \cite{kikuchi2016fundamentals}. This latter problem, phase reference frame sharing considered in the context of  CV-QKD, will be the main focus of this article . 

The problem of sharing a reference frame is specific in the sense that reference frame information constitutes \emph{unspeakable information}, that can only be shared through physical carriers exchanged between emitter and receiver \cite{BartlettRudolphSpekkens2007}. On the other hand, it is important to emphasize that although quantum mechanics gives a precise framework to formulate the question of reference frame sharing, in relation with quantum metrology \cite{BartlettRudolphSpekkens2007}, this question can be solved ``classically'', using macroscopic signals to exchange reference frame information. The type of questions related to phase reference sharing is not whether it is possible, but whether it can be achieved given resource constraints, dictated by the hardware resources and by the characteristics of the channel, such as losses and noise.  In line with the recent work on LLO CV-QKD  \cite{LLOexp1,LLOexp2,LLOexp3}, we will focus on in this article on the issue of jointly performing, with the same hardware, phase reference sharing and CV-QKD.

Another question related to  reference frame sharing in quantum communications consist in  performing ``referenceless'' quantum communication in which  quantum information is encoded so that it can be recovered ``without reference frame'' at the receiver, under some assumptions about the channel such as collective noise.  Such approach can for example be used with polarization encoding, when emitter and receiver spatial reference frame are slowly rotating, by encoding quantum information over noiseless subspaces  \cite{LaflammeSpekkens2003,BoileauLaflamme2006,ScaraniObrien2010}. Referenceless quantum communications can be seen as a specific approach to perform reference frame sharing. This approach however requires to encode information over entangled quantum states, and cannot be easily used to design practical optical encodings for CV-QKD. We will therefore not consider the referenceless protocols in this article.\\

\textbf{Phase reference sharing in classical and quantum coherent communication.} Coherent communication systems have the advantage of offering higher sensitivity (information per photon) than systems based on direct detections (for example On-Off-Keying modulation, where the information is encoded solely in intensity), and classical coherent systems are gradually becoming more and more used in modern classical optical networks, especially in core networks, over long-distance segments.
Phase reference sharing is an important requirement in coherent communication systems, in order to correct the phase drift between the phase of the emitter laser and the phase of the local oscillator laser, placed at the receiver side. The generic objective is essentially to solve this phase reference sharing problem with minimal resource overhead and minimum penalty on the associated communication protocol.  An essential point is to notice that the constraints and thus the solutions that can be adopted in the classical and in the quantum cases to solve the phase reference sharing issue significantly differ.

Classical coherent detectors are designed to detect intense light pulses, typically coherent states containing a very large number of photons, while coherent detectors used in quantum communications must typically be operated in the shot-noise regime, i.e with electronic noise significantly below the signal variance associated with the detection of one photon. This limits the intensity that can be handled by shot-noise limited coherent detectors before saturation \cite{SaturationAttackHao}. While analogic phase lock loops where used until the 70's in classical coherent optical systems to solve the issue of phase locking, suffering however from phase lock loop bandwidth limitations, the advent of GHz-clocked electronics and fast digital signal processing now allows to recover both signal information and phase reference information from discrete modulation, such as binary phase shift keying (BPSK) or higher order modulations \cite{kikuchi2016fundamentals}. Such phase recovering techniques, used for classical coherent communication systems, that require a high number of photons at reception and high-speed modulations/detections, cannot be directly applied to perform quantum coherent communications. This makes the problem of phase reference sharing in quantum communications more constrained, and requires specific approaches.

Since the problem of phase reference sharing can be solved by sending classical reference pulses, one simple approach to the problem, in the context of quantum coherent communication is indeed to use an external classical (intense) phase reference sharing scheme. Such ``classical'' method is always possible but will typically requires the use of two separate detectors, one shot-noise limited detector for (weak) quantum signals and a second detector, with a large linearity range, to detect classical phase reference signals. This classical method hence implies not only techniques for multiplexing and demultiplexing reference and quantum signals, but basically to deploy two separate detection hardwares.

It is natural to seek how one can lift the extra hardware requirement of the ``classical'' method in order to jointly perform  phase reference sharing and quantum communication with the same hardware. This question has been addressed in recent works aiming at demonstrating CV-QKD operation with a local local oscillator (LLO) \cite{LLOexp1,LLOexp2,LLOexp3}, however with performance limitations and constraints on the hardware. \\

\textbf{Contributions of this work.} We identify and discuss the existing approaches to the phase reference sharing problem for LLO CV-QKD. Recent works \cite{LLOexp1,LLOexp2,LLOexp3} all rely on time-multiplexed quantum signals pulses with reference pulses in order to jointly perform phase recovery and quantum communication. In this work, we introduce new elements in the standard noise model of CV-QKD analysis, considering new practical constraints imposed by the simultaneous quantum signal and phase reference transmission of LLO-based CV-QKD. In particular, the amplitude modulator (AM) dynamics and the linearity range of Bob's detector are studied and we show that the AM dynamics is a key parameter in order to compare performance of realistic implementations of LLO-based CV-QKD. As a contribution, our resulting noise model is a refined framework for realistic CV-QKD analysis, including LLO regimes. Based on this comprehensive model, we show that there exist fundamental and practical limitations in the phase noise tolerance of the designs introduced in \cite{LLOexp1,LLOexp2,LLOexp3}, that we designate as LLO-sequential. 

In order to go beyond that phase noise limit, we introduce the idea of self-coherence in phase reference sharing for CV-QKD implementations based on a local local oscillator. Self-coherent designs consist in ensuring the phase coherence between pairs of quantum signal and phase reference pulses by deriving both of them from the same optical wavefront at emission. This allows to perform relative phase recovery schemes with better sensitivity than in the LLO-sequential design. In particular, we propose a design, called LLO-delayline, implementing a self-coherent phase sharing design. It ensures the self-coherence using a balanced delay line interferometer split between emitter and receiver sides. We analyze how self-coherence is obtained and study the performance reachable with this design, demonstrating that they exhibit a much stronger resilience to high phase noise than the LLO-sequential design under realistic experimental parameters. While previous experimental proposals of LLO CV-QKD are limited to slowly varying reference frames regimes (ie. based on very stable lasers or high repetition rates), our newly introduced design allows phase reference sharing resilient to high phase noise regimes, using the idea of self-coherence. 

A second self-coherent design, referred to as LLO-displacement, relies on an original multiplexing allowing to transmit both the quantum signal and the reference pulse within each optical pulse. The simultaneous transmission of quantum signal and phase reference can be seen as an original cryptographic primitive, considered in \citep{Qi2016}, that can be used with different modulation schemes. In particular, this allows to optimize the resources $-$ in terms of required hardware and repetition rate $-$ in LLO-based CV-QKD experiments. We also emphasize that an important advantage of our LLO-displacement design is its experimental simplicity as we show that the multiplexing can be perform numerically on Alice's variables. As such, no specific hardware devices are required. We study the theoretical performance of such design and exhibit its limitations.

In Sec.~\ref{Sec: Previous Work}, we review the existing implementations of phase reference sharing CV-QKD. In Sec.~\ref{Sec: CVQKD}, we introduce the CV-QKD model. In particular, the phase reference sharing issue in CV-QKD is formally introduced and discussed and we also introduce our comprehensive noise model. In Sec.~\ref{Sec: Towards self-coherence}, we highlight practical limitations of existing local local oscillator based CV-QKD and introduce the idea of self-coherence for reference sharing in CV-QKD. In Sec.~\ref{Sec: LLO-delayline} and Sec.~\ref{Sec: LLO-displacement}, we respectively introduce the LLO-delayline and LLO-displacement designs and study their performance. Conclusion and perspectives are presented in Sec.~\ref{Sec: Conclusion}.

\section{Implementing phase reference sharing in CV-QKD: previous work}
\label{Sec: Previous Work}

The procedure used for phase reference sharing in quantum coherent communication is often not tackled explicitly in experiments. As mentioned in the introduction, this follows from the idea that this question can in principle be solved independently of the quantum communication protocol itself, with classical techniques. This sometimes motivates to only perform phase reference sharing by placing emitter and receiver in the same location and using locally one single laser source both for quantum signal preparation and as local oscillator. Such proof of principle  implementations have been used in early CV-QKD demonstrations \cite{grosshans2003high} and, more recently, in experimental demonstrations of measurement device independent (MDI) CVQKD \cite{pirandola2015high}.

In more realistic experimental demonstrations, emitter and receiver must be placed in distant locations and some specific design must be used in order to obtain a local oscillator, at the receiver side, phase locked with the emitter laser. The simplest experimental approach is actually to use the laser at the emitter side to generate the local oscillator, and to send it to the receiving side, using adapted multiplexing schemes. This procedure is called the transmitted LO design (TLO) and we will review its principle and its limitations. As we will see, TLO suffers from a fundamental weakness in the cryptographic context of CV-QKD, due to the security loophole associated with LO manipulation as it propagates on a public channel. This has lead to implement CV-QKD with a true local local oscillator (LLO), and we will review the recent work in this direction.

\subsection{The transmitted local oscillator (TLO) design}
\label{Sec: TLO}

\begin{figure*}
\centering
\includegraphics[scale=0.75,trim = 6.5cm 7cm 8.5cm 7.5cm, clip]{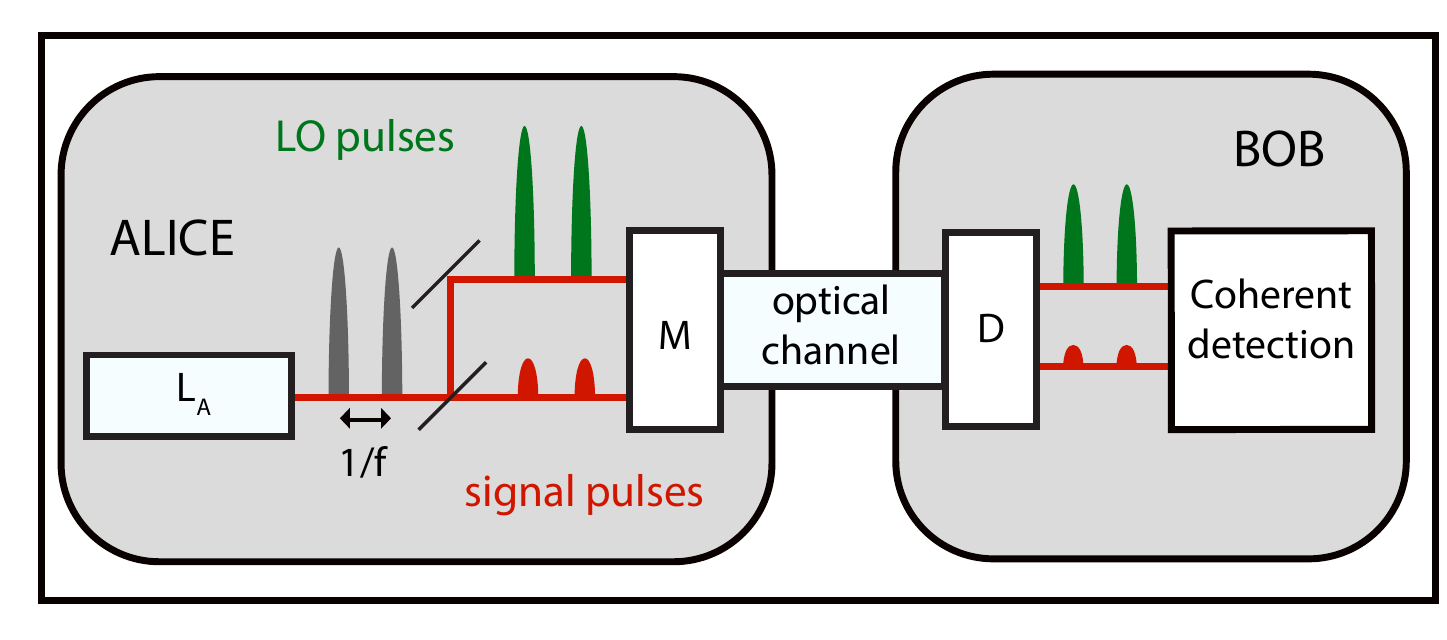}
\caption{\textit{(color) Transmitted local oscillator (TLO) design. In the TLO design, the phase reference (green pulse) and the quantum signal (red pulse) are derived from the same optical pulse and sent from Alice to Bob using multiplexing/demultiplexing (M/D) techniques.}}
\label{Scheme: TLO}
\end{figure*}

In most implementations of CV-QKD performed so far \cite{ExpCVQKD-chinois2,ExpCVQKD-chinois,ExpCVQKD-Sequrenet,ExpCVQKD-Qi}, the phase reference is directly transmitted from Alice to Bob through the optical channel as a bright optical pulse multiplexed in time and polarization with each quantum signal pulse and is used as the LO pulse at reception. Such implementation is detailed in Fig.~\ref{Scheme: TLO} and is referred to as the Transmitted LO (TLO) design. The main advantage of this scheme is the guarantee, by design, of a stable relative phase between quantum signal and LO at reception by producing both of them from a single laser L$_A$ placed at Alice's side. An interferometric setup, based on polarization delay-line interferometers, is used to multiplex (M) and demultiplex (D) the quantum signal and the LO, hence ensuring a low relative phase noise at reception. The only limitation in terms of tolerable phase noise is that the phase of the laser can be considered as stable over the duration of a single optical pulse, resulting in $\Delta\nu /f \sim 10$, where $\Delta \nu$ is the spectral linewidth of the laser, $f$ is the repetition rate and we assume a typical pulse duration of $0.1/f$. Despite it is the most implemented GMCS protocol, security weaknesses of such implementations have however been demonstrated in practice by manipulating the LO intensity \cite{grosshans2007, ma2014,CalibAttacks} or wavelength \cite{weedbrook2013} on the quantum channel.

Furthermore, based on a coherent detection at reception, such protocols rely on the use of a bright LO at reception (around $10^8$ photons per pulse at reception are required to ensure that the coherent detection can be operated with low electronic to shot noise ratio in \cite{ExpCVQKD-Sequrenet}). For long distance or high speed (where the pulse duration is short), the requirements in terms of launch power at emission creates practical issues. Because of limited power of lasers as well as Brillouin effect and non-linear effects in optical fibers \cite{brillouin1, brillouin2,nonlinearoptics}, there is a typical limit of few tens of milliwatts on the launched power of each involved laser for CV-QKD purposes. In the TLO design, this limit is a major limitation of the LO intensity at reception, especially for long distances. This  will in particular limit the possibility of using the TLO design on shared optical fibers at long distance and high-rate operation, i.e. situations where the requirements on LO power at emission would be extremely large.

\subsection{The local local oscillator (LLO) sequential design}
\label{Sec: LLO-seq}

In order to lift the important limitations (both theoretical and practical) of CV-QKD implementations relying on the TLO design, a new CV-QKD method relying on a ``local local oscillator'' (LLO) has recently been independently introduced in \cite{LLOexp1,LLOexp2,LLOexp3}. This method, implementing the Gaussian modulated coherent state protocol, consists in using a second laser at Bob's side in order to produce local LO pulses for coherent detections. One crucial advantage of implementing CV-QKD in a LLO configuration is to close, by design, any potential security loophole linked to the possibility of manipulating the LO as it propagates on the public optical channel between Alice and Bob. Implementing LLO CV-QKD allows on the other hand to ensure by design that the LO is fully trusted, and in particular that the LO amplitude (that requires careful calibration) cannot be manipulated. Another important advantage of LLO CV-QKD stems from the fact that in this configuration, repetition rate and distance do not affect the LO intensity at detection. A LO power sufficient to ensure high electronic to shot noise ratio may thus be obtained, independently of the propagation distance.

\begin{figure*}
\centering
\includegraphics[scale=0.8,trim = 6.5cm 7cm 8.2cm 8cm, clip]{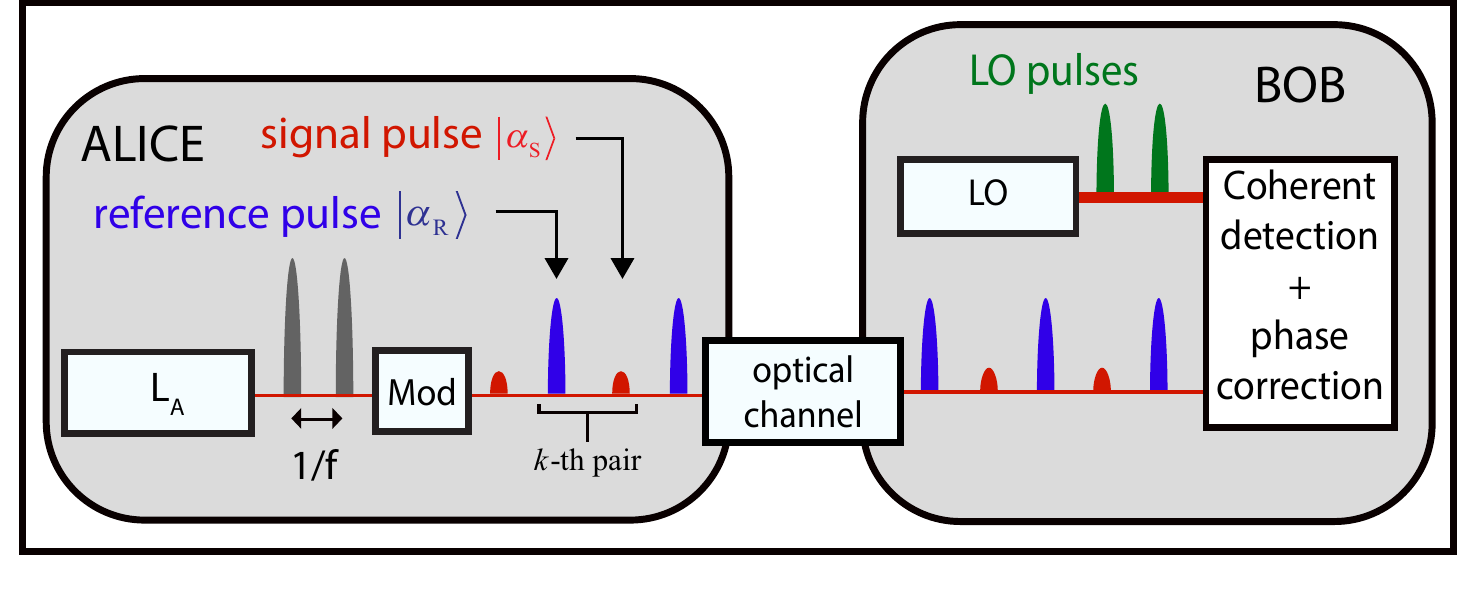}
\caption{\textit{(color) Local local oscillator (LLO) sequential design. In the LLO-sequential design, Alice sequentially sends weak quantum signal (red pulse) and bright phase reference (blue pulse) pulses. At reception, Bob performs consecutive coherent detections of each pulse received using is own LO pulses (green pulse).}}
\label{Scheme: LLO-Sequential}
\end{figure*}

Implementing CV-QKD in the LLO configuration however comes with new experimental challenges. The main issue in LLO-based CV-QKD is to be able to perform CV-QKD despite the potentially important drift of the relative phase between Alice's emitter laser L$_A$ and Bob's local oscillator laser L$_B$, see Fig.~\ref{Scheme: LLO-Sequential}. The relative phase at reception is, in the case of LLO-based CV-QKD, the relative phase between the two free-running lasers L$_A$ and L$_B$. As such, Bob's raw measurement outcomes are \emph{a priori} decorrelated from Alice's quadratures and a phase correction process has to be performed in order to allow secret key generation. The goal of the phase reference sharing in the context of LLO CV-QKD is then to ensure a low enough phase noise so that the excess noise is significantly below the threshold imposed by security proofs \cite{ReviewCVQKD}. 

Recent works \cite{LLOexp1,LLOexp2,LLOexp3} have demonstrated the possibility of implementing the GMCS protocol using a local local oscillator, by introducing an experimental design, depicted on Fig.~\ref{Scheme: LLO-Sequential}, that we will call LLO-sequential. In the LLO-sequential design, Alice sequentially sends, at a repetition rate $f/2$, consecutive pairs $(|\alpha_\mathrm{S} \rangle, | \alpha_\mathrm{R} \rangle)$ of coherent states where $|\alpha_\mathrm{S} \rangle$ is a GMCS quantum signal pulse and $|\alpha_\mathrm{R } \rangle = | E_\mathrm{R} \rangle$ is a phase reference pulse with a fixed phase set to $0$ and an amplitude $E_\mathrm{R}$. Phase reference pulses are relatively bright pulses compared to the signal and have a fixed phase in Alice's phase reference frame, that is publicly known so that it carries information on Alice's reference frame. At reception, Bob performs sequential coherent detections of quantum signal and phase reference, using a single detector, operated with a ``local local oscillator'', placed at Bob. Bob can thus estimate the relative phase using the phase reference pulse and a phase correction can be performed on Alice and Bob's signal data in order to generate secret key.

In \cite{LLOexp2} a 250 kHz-clocked proof-of-principle experiment of the LLO-sequential designed is performed, however with only one single laser playing the role of both emitter and LO, and two consecutive uses of a homodyne detector used to emulate a heterodyne measurement. In \cite{LLOexp1}, another proof-of-principle experiment with two lasers and Alice and Bob connected by a 25 km optical fiber is performed with a 50 MHz-clocked system. The authors demonstrate that phase correction can be implemented with a residual excess noise compatible with CV-QKD security threshold. Joint operation of CV-QKD (requiring weak quantum signals) together with the phase correction mechanism (requiring bright phase reference pulses) was studied through a simulation, which left aside the question of the hardware requirements for both CV-QKD and phase reference sharing. \cite{LLOexp3} provides a whole experimental demonstration of an implementation of LLO-sequential CV-QKD over 25 km, with a 100 MHz-clocked system, and a 1 GHz-bandwidth shot-noise limited homodyne detection. We should however emphasize that these strong experimental performances have been hove we obtained expensive hardware, namely two low phase noise ECL lasers, as emitter and LO, and an amplitude modulator with 60 dB of dynamics. Such hardware is typically not available in standard telecom environment and the issue of considering LLO CV-QKD implementation with realistic hardware  should be addressed in order to study the ability to be ubiquitously integrate CV-QKD within modern optical networks. 

\section{CV-QKD: protocol and noise model} 
\label{Sec: CVQKD}

Different CV-QKD protocols have been proposed so far including protocols based on squeezed states of the electromagnetic field \cite{GarciaCerf2009} or on discrete modulations of coherent states \cite{GrangierLeverrier2009,BecirWahiddin2010}. However, squeezed states are experimentally challenging to produce and security analysis for discrete modulation CV-QKD are less advanced. Due to its experimental convenience \cite{ExpCVQKD-Sequrenet,ExpCVQKD-Qi,ExpCVQKD-chinois} and its good security analysis understanding \cite{LeverrierThesis,GarciaPatronThesis}, the Gaussian-modulated coherent states (GMCS) protocol is the most implemented CV-QKD protocol and has reached the step of commercialization \citep{sequrenet}. 

It is therefore natural to consider GMCS CV-QKD in order to study CV-QKD with a local local oscillator and to perform early experiment, as it has been the case in  \cite{LLOexp1,LLOexp2,LLOexp3}. We then also focus our analysis on GMCS CV-QKD and introduce new elements in the noise model in order to account for the important constraints that drive the performance of CV-QKD in the regime of a local local oscillator. This in particular allows us to discuss the limitations of  LLO-sequential design when implemented with realistic hardwares.

\subsection{GMCS protocol and secret key rate}

In the Gaussian-modulated coherent states protocol, Alice encodes classical Gaussian variables $(x_A,p_A)$ on the mean values of the two conjugate quadratures of coherent states $|\alpha \rangle = |x_A,p_A \rangle$. Coherent states are then sent to Bob through an insecure channel controlled by an eavesdropper Eve. At reception, Bob performs a coherent detection of either one quadrature (homodyne detection) or both quadratures (heterodyne detection) of the received pulse and calculates estimators $(x_B,p_B)$ of Alice's variables. As Eve's optimal attacks are Gaussian \cite{GarciaPatronThesis,LeverrierThesis}, we can model the logical channels between Alice and Bob's data as additive white Gaussian noise  channels \cite{FossierThesis}:
\begin{eqnarray}
\begin{array}{ccc}
x_B & = & \sqrt{\frac{G}{\delta_\mathrm{det}}} \cdot \left( x_A + x_0 + x_c \right)\\
p_B & = & \sqrt{\frac{G}{\delta_\mathrm{det}}} \cdot \left( p_A + p_0 + p_c \right)
\end{array}
\label{Eq: Gaussian channel model}
\end{eqnarray}

where $(x_c,p_c)$ is the total noise of the channel and we note $(\chi_x,\chi_p)$ its variance. In Eq.~\ref{Eq: Gaussian channel model}, $G$ is the total intensity transmission of the channel, $\delta_\mathrm{det}$ stands for the detection used at reception ($\delta_\mathrm{det}=1$ for a homodyne detection and $\delta_\mathrm{det}=2$ for a heterodyne detection), $x_0$ and $p_0$ are Gaussian variables of variance $N_0$ modelling the shot noise quadratures. In general \cite{LeverrierThesis}, it is assumed that the channel noise is symmetric and $\chi=\chi_x=\chi_p$ where the variance $\chi$ is referred to Alice's input. The variance $\chi$ of the total noise can be expressed as \cite{FossierThesis,GarciaPatronThesis}:
\begin{eqnarray}
\chi & = & \frac{\delta_\mathrm{det}-G}{G} + \xi
\label{Eq: Total noise}
\end{eqnarray}
where the first term is the loss-induced vacuum noise and $\xi$ is the overall excess noise variance of the channel referred to Alice's input. Thereby, using Eq.~\ref{Eq: Gaussian channel model} and Eq.~\ref{Eq: Total noise}, we can see that the Gaussian channel between Alice and Bob is fully characterized by the two parameters $G$ and $\xi$. In a real-world experiment, Alice and Bob can estimate $G$ and $\xi$ from the correlations between their respective variables by revealing a fraction of their data and are then able to characterize the propagation channel and generate secret key. We discuss and model the different contributions to the excess noise $\xi$ in practical CV-QKD in the next paragraph. Finally, the secret key rate available to Alice and Bob in the reverse reconciliation scheme can be expressed as \cite{FossierThesis,GarciaPatronThesis}:
\begin{eqnarray}
k & = & \beta\cdot I_{AB} - Q_{BE}
\end{eqnarray}

where $0 \leq \beta \leq 1$ is the reconciliation efficiency, $I_{AB}$ is the mutual information between Alice and Bob's classical variables and $Q_{BE}$ stands for Eve's maximal accessible information on Bob's measurements, capturing assumptions on Eve's behaviour \cite{GarciaPatronThesis}. In this work, we restrict the security analysis to individual attacks \cite{FossierThesis,GarciaPatronThesis} and $Q_{BE}$ is then the classical information $I_{BE}$ between Bob's measurements and Eve's data. In \cite{ExpCVQKD-Sequrenet,LLOexp1}, it is assumed that Eve does not have access to Bob's electronics. In this work however, we use the stronger security model of \cite{LLOexp2} assuming that Eve is able to control the noise of Bob's detector.

\subsection{Noise model}
\label{Sec: Noise model} 

Implementing CV-QKD with a local local oscillator comes with new challenges. The main challenge is related to the fact that Alice and Bob must use a procedure to compensate efficiently the phase drift between two different lasers, used respectively as emitter and local oscillator, in order to be able to perform CV-QKD with a tolerable noise level. Another more specific aspect of the challenge is related to the objective targeted in this paper: propose and study practical implementation schemes for LLO CV-QKD with shared and affordable hardware: this leads to consider practical limitations that had been previously overlooked, and allows to study resource trade-off.\\

\paragraph{Relative phase noise.} In LLO-based CV-QKD, the main challenge is to create a reliable phase reference between emitter and receiver because the relative phase drift between the two involved lasers may fully decorrelate Alice's variables and Bob's measurements thus preventing any secret key rate generation. We define the \emph{signal relative phase} $\theta_\mathrm{S}$ as the phase difference between the LO pulse $|\alpha_\mathrm{LO } \rangle$ and the signal pulse $|\alpha_\mathrm{S} \rangle$ at reception:
\begin{eqnarray}
\theta_\mathrm{S} & = & \varphi_\mathrm{LO}-\varphi_\mathrm{S}
\label{Eq: Relative Phase}
\end{eqnarray}
where $\varphi_\mathrm{S}$ is the signal phase and $\varphi_\mathrm{LO}$ is the phase of the local oscillator at reception. Using the notations of Eq.~\ref{Eq: Gaussian channel model} and in presence of a relative phase $\theta_\mathrm{S}$, we can write Bob's measurement outcomes, when performing an heterodyne as:
\begin{eqnarray}
\begin{pmatrix} x_B \\ p_B \\ \end{pmatrix} & = & \sqrt{\frac{G}{\delta_\mathrm{2}}} \cdot \left[ 
\begin{pmatrix}\cos \theta_\mathrm{S} & \sin \theta_\mathrm{S}\\ 
-\sin\theta_\mathrm{S} & \cos \theta_\mathrm{S}\\ \end{pmatrix}  \cdot 
\begin{pmatrix} x_A \\ p_A \\ \end{pmatrix} + 
\begin{pmatrix} x_0 + x_c  \\ p_0 + p_c \\ \end{pmatrix}  \right]
\label{Eq: Bob's measurement outcomes}
\end{eqnarray}

where $x_c$ and $p_c$ capture all excess noise sources but the phase noise. The relative phase $\theta_\mathrm{S}$  acts as the selector of the measured quadrature.
In the TLO design, the relative phase $\theta_\mathrm{S}$ is, by design, always close to $0$. However, in the case of two free-running lasers, $\theta_\mathrm{S}$ depends on the relative phase $\theta$ between the two lasers. Assuming that the two lasers L$_A$ and L$_B$ are centered around the same optical frequency and have spectral linewidths 
 $\Delta \nu_A$ and $\Delta \nu_B$, we can model \cite{Schulze2005,YarivYeh2006,Bittner2010} the relative phase $\theta=\varphi_B-\varphi_A$ ($\varphi_A$ and $\varphi_B$ are respectively the phase of L$_A$ and L$_B$) as a Gaussian stochastic process $\{\theta_t\}_t$ characterized by the variance of the drift between two times $t_i$ and $t_{i+1}$:
\begin{eqnarray}
\mathrm{var}\left( \theta_{i+1} | \theta_i \right) & = & 2\pi \cdot (\Delta \nu_A + \Delta \nu_B) \cdot |t_{i+1} - t_{i}|
\label{Eq: Laser Phase Drift Model}
\end{eqnarray}
where $\theta_{i}$ and $\theta_{i+1}$ correspond to the relative phase at consecutive times $t_i$ and $t_{i+1}$. 

We can see from Eq.~\ref{Eq: Bob's measurement outcomes} that this implies a decorrelation between Alice's data and Bob's measurements which can be seen as a contribution, noted $\xi_\mathrm{phase}$, to the excess noise $\xi$. The principle of phase reference sharing schemes considered in the article consists in using a reference pulse to build an estimate $\hat{\theta}_\mathrm{S}$ of the actual relative phase $\theta_\mathrm{S}$ of the signal (relative means relative with respect to local oscillator), and to apply a phase correction $- \hat{\theta}_\mathrm{S}$ on the signal, in order to compensate for the phase drift. In a reverse reconciliation scheme, this correction has to be performed on Alice's data as a rotation of her data:
\begin{eqnarray}
\begin{pmatrix} \tilde{x_A} \\ \tilde{p_A} \\ \end{pmatrix} & = & \begin{pmatrix} \cos \hat{\theta}_\mathrm{S} & \sin \hat{\theta}_\mathrm{S}\\ 
-\sin \hat{\theta}_\mathrm{S} & \cos \hat{\theta}_\mathrm{S}\\ \end{pmatrix} \cdot \begin{pmatrix} x_A \\ p_A \\ \end{pmatrix}
\label{Eq: Alice phase correction}
\end{eqnarray}

We can show from Eq.~\ref{Eq: Total noise}, \ref{Eq: Bob's measurement outcomes}, \ref{Eq: Alice phase correction} that the remaining excess noise $\xi_\mathrm{phase}$  due to phase noise (after correction) depends on the modulation format and is, in general, not symmetric on the two quadratures . However, in the case of the GMCS protocol (with modulation variance $V_A$) and assuming that the remaining phase noise $\theta_\mathrm{S} - \hat{\theta}_\mathrm{S}$ after correction is Gaussian, the phase noise $\xi_\mathrm{phase}$ can then be written as:
\begin{eqnarray}
\xi_\mathrm{phase} & = &  2V_A \cdot \left(1 - e^{-V_\mathrm{est}/2} \right)
\label{Eq: GMCS Phase noise}
\end{eqnarray}
where we define the variance $V_\mathrm{est}$ of the \textit{remaining phase noise} (after reference quadrature measurement, relative phase estimation and correction) as:
\begin{eqnarray}
V_\mathrm{est} & \hat{=} & \mathrm{var}\left(\theta_\mathrm{S} - \hat{\theta}_\mathrm{S}\right)
\label{Eq: General phase estimation variance}
\end{eqnarray}
Eq.~\ref{Eq: GMCS Phase noise}  (derived in Annex.~\ref{Annex: Phase excess noise}) is a generalization of the phase noise expression given in \cite{LLOexp2,Qi2016} for the case of small phase noise.

An important challenge to perform LLO-based CV-QKD is therefore to calculate a precise estimator $\hat{\theta}_\mathrm{S}$ of the relative phase $\theta_\mathrm{S}$ in order to minimize $V_\mathrm{est}$ and thus $\xi_\mathrm{phase}$. 
The general scheme for phase reference sharing design in LLO  CV-QKD can be modeled in the following way: Alice generates two coherent states, a quantum signal pulse $|\alpha_\mathrm{S} \rangle$ and a reference pulse $|\alpha_\mathrm{R} \rangle$ and sends them on the optical channel using some multiplexing scheme. At reception, Bob performs demultiplexing, and uses the received reference pulse to derive an estimate $\hat{\theta}_\mathrm{R}$ the relative phase $\theta_\mathrm{R}$ between the reference pulse and 
the local oscillator at reception.  The phase sharing designs (cf. sections \ref{Sec: Towards self-coherence}, \ref{Sec: LLO-delayline}, \ref{Sec: LLO-displacement}) give guarantees that the relative phase of the reference pulse is close to the relative phase of the quantum signal, i.e.  that $\theta_\mathrm{R} \approx \theta_\mathrm{R}$. Therefore,  the estimated value $\hat{\theta}_\mathrm{R}$  can be used to approximate and then correct the relative phase of the signal $\theta_\mathrm{S}$.

\begin{figure*}
\begin{center}
\includegraphics[scale=0.8,trim = 1cm 9cm 5cm 5cm, clip]{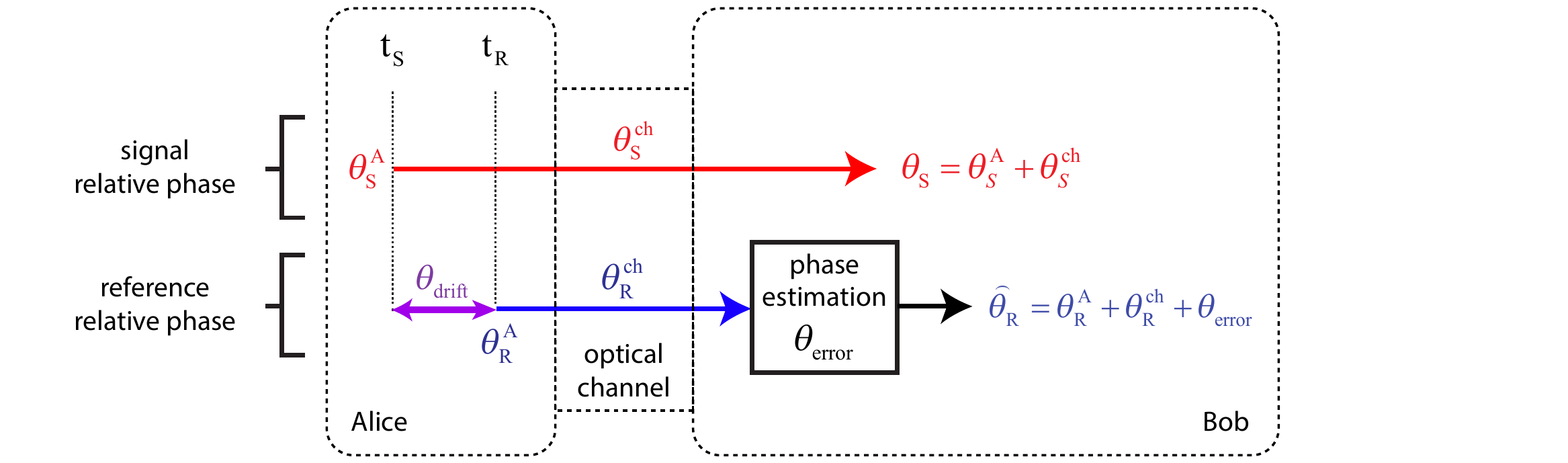}
\caption{\textit{(color) Schematic representation of a general relative phase estimation process. The relative phase $\theta_\mathrm{S}$ at reception (red), which is defined with respect to the $LO$ phase (Eq.~\ref{Eq: Relative Phase}), is estimated at reception with the estimator $\hat{\theta}_\mathrm{R}$ inferred from specific reference phase information evaluation (blue).}}
\label{Figure: General phase estimation process}
\end{center}
\end{figure*}

A general picture of the phase estimation process, and of the sources of deviations, is depicted in Fig.~\ref{Figure: General phase estimation process}. We can express the quantum signal relative phase $\theta_\mathrm{S}$ (with respect to local oscillator) as the sum of the relative phase $\theta_\mathrm{S}^\mathrm{A}$ at emission  and the phase $\theta_\mathrm{ch}^\mathrm{S}$ accumulated by the coherent state $|\alpha_\mathrm{S} \rangle$ on the optical channel: 
\begin{eqnarray}
\theta_\mathrm{S} & = & \theta_\mathrm{S}^\mathrm{A} + \theta_\mathrm{S}^\mathrm{ch}
\label{Eq: Signal relative phase}
\end{eqnarray}

Similarly and using the same notations for the reference pulse $|\alpha_\mathrm{R} \rangle$, we can express the relative phase $\theta_\mathrm{R}$ at reception of a reference pulse $|\alpha_\mathrm{R} \rangle$ as:
\begin{eqnarray}
\theta_\mathrm{R} & = & \theta_\mathrm{R}^\mathrm{A} + \theta_\mathrm{R}^\mathrm{ch}
\label{Eq: Ref relative phase}
\end{eqnarray}

At reception, Bob measures both quadratures of $|\alpha_\mathrm{R} \rangle$ using a heterodyne detection (in the remaining of the paper, we only consider heterodyne detections at reception with $\delta_\mathrm{det}=2$). The estimator $\hat{\theta}_\mathrm{R}$ of $\theta_\mathrm{R}$ can be calculated from the heterodyne measurement outcomes $x_B^{\mathrm{(R)}}$ and $p_B^{\mathrm{(R)}}$ as:
\begin{eqnarray}
\hat{\theta}_\mathrm{R} = \tan^{-1}\left(\frac{p_B^{\mathrm{(R)}}}{x_B^{\mathrm{(R)}}}\right)
\end{eqnarray}
Due to the fundamental shot noise and to the experimental noise on the heterodyne detection, $\hat{\theta}_\mathrm{R}$ differs from$\hat{\theta}_\mathrm{R}$ by an error $\theta_\mathrm{error}$ characterized by its variance:
\begin{eqnarray}
V_\mathrm{error} & \hat{=} & \mathrm{var} \left(\hat{\theta}_\mathrm{R} - {\theta}_\mathrm{R} \right)
\end{eqnarray}
We can show that, in the case of a reference pulse of the form $|\alpha_\mathrm{R} \rangle = |E_\mathrm{R} \rangle$:
\begin{eqnarray}
V_\mathrm{error} & = & \frac{\chi + 1}{E_\mathrm{R}^2}
\label{Eq: Single pulse phase estimation}
\end{eqnarray}

where $E_\mathrm{R} = |\alpha_\mathrm{R}|$ is the amplitude of the reference pulse and $\chi$ is defined in Eq.~\ref{Eq: Total noise}.
Finally, Bob uses the relative phase estimate  $\hat{\theta}_\mathrm{S}$ to apply a phase correction $- \hat{\theta}_\mathrm{R}$ to the quantum signal. The overall process is schematically represented in Fig.~\ref{Figure: General phase estimation process}.
It results, after phase correction, to a remaining phase noise $V_\mathrm{est}=\mathrm{var}\left(\hat{\theta}_\mathrm{R} - \theta_\mathrm{S}\right)$ which can be expressed using Eq.~\ref{Eq: Signal relative phase} and  Eq.~\ref{Eq: Ref relative phase} as:  

\begin{eqnarray}
V_\mathrm{est} & = & V_\mathrm{error} + V_\mathrm{drift}  + V_\mathrm{channel}
\label{Eq: Total remaining phase noise}
\end{eqnarray}

where:
\begin{eqnarray}
V_\mathrm{drift} & \hat{=} & \mathrm{var}\left( \theta_\mathrm{R}^\mathrm{A} - \theta_\mathrm{S}^\mathrm{A}\right) \\
V_\mathrm{channel} & \hat{=} & \mathrm{var}\left( \theta_\mathrm{R}^\mathrm{ch} - \theta_\mathrm{S}^\mathrm{ch}\right)
\label{Eq: Vchannel}
\end{eqnarray}

The term $V_\mathrm{drift}$ corresponds to the variance of the relative phase drift 
$ \theta_\mathrm{drift}=\theta_\mathrm{R}^\mathrm{A} - \theta_\mathrm{S}^\mathrm{A}$
between the two free-running lasers L$_A$ and L$_B$ between time $t_\mathrm{S}$ at which $|\alpha_\mathrm{S} \rangle$ is emitted and time $t_\mathrm{R}$ at which $|\alpha_\mathrm{R}\rangle$ is emitted.  From  Eq.~\ref{Eq: Laser Phase Drift Model}, we can express the phase noise due to laser phase drift between times $t_\mathrm{S}$ and $t_\mathrm{R}$ as:
\begin{eqnarray}
V_\mathrm{drift}  = & 2\pi \cdot (\Delta \nu_A + \Delta \nu_B) \cdot |t_\mathrm{R} - t_\mathrm{S}|
\label{Eq: Relative phase drift R-S}
\end{eqnarray}

We can observe that the time delay between signal and phase reference emissions implies a decorrelation between the corresponding relative phases and, thus, introduce a noise on the phase estimation process. This leads to the main limitation of the LLO-sequential approach as explained in next section.\\

The term $V_\mathrm{channel}$ corresponds to the relative phase drift due to the difference of the phase accumulated by $|\alpha_\mathrm{S} \rangle$ and $|\alpha_\mathrm{R} \rangle$ during propagation. In practice, we assume that this term is dominated by the difference between the optical path lengths of $|\alpha_\mathrm{S} \rangle$ and $|\alpha_\mathrm{R} \rangle$. 
 
In the remaining of this article we will study and discuss the performance of existing as well as newly introduced LLO based CV-QKD designs, relying on different relative phase sharing designs. For each of these designs, we will explicit the expressions of the different contributions to the remaining phase noise $V_\mathrm{est}$ of Eq.~\ref{Eq: Total remaining phase noise}.\\

\paragraph{Electronic to shot noise ratio.} Intrinsic electronic noise of Bob's detector induces a noise of variance $v_\mathrm{elec}$ in shot noise unit (SNU) on Bob's quadrature measurements. As the shot noise value is linear with the LO intensity, it is however possible to reduce the effective electronic to shot noise ratio $\xi_\mathrm{elec}$ by increasing the LO intensity. We model $\xi_\mathrm{elec}$, referred to Alice's input, as:
\begin{eqnarray}
\xi_\mathrm{elec} & = & \frac{\delta_\mathrm{det}}{G} \cdot \frac{E^2_\mathrm{LO,cal} \cdot v_\mathrm{elec}}{E_\mathrm{LO}^2}
\label{Eq: electronic noise}
\end{eqnarray}
where $E_\mathrm{LO,cal}^2$ is the photon number in the LO at which the electronic noise is $v_\mathrm{elec}$ at Bob side and $E_\mathrm{LO}^2$ is the actual photon number per LO pulse at reception. Furthermore, we consider that Eve is able to manipulate the electronic noise which corresponds to a strong security scenario \cite{LLOexp2}.\\ 

\paragraph{Amplitude modulator finite dynamics.} Amplitude modulators efficiency are limited by their dynamics restricting the range of the achievable transmission coefficient. Recent works \cite{LLOexp1,LLOexp2,LLOexp3} have proposed to conjointly communicate weak quantum signals and relatively bright reference pulses using a single experimental setup and, in particular, a single amplitude modulator (AM). This directly adresses the issue of the AM dynamics at emission, limitating the maximal amplitude that Alice can output and introducing a leakage on the amplitude modulated. The ratio between the maximal and minimal amplitudes $E_{max}$ and $E_{min}$ that Alice can output is characterized by the dynamics $\mathrm{dyn}_\mathrm{dB}$ of the AM defined as:
\begin{eqnarray}
\mathrm{dyn}_\mathrm{dB} & = & 10 \cdot \log_{10} \left(\frac{E_{max}^2}{E_{min}^2} \right)
\end{eqnarray}
From this equation, one can model Alice's modulator imperfection as an amplitude leakage on each optical pulse resulting in an excess noise which can be approximated as:
\begin{eqnarray}
\xi_\mathrm{AM} & = & E_\mathrm{max}^2\cdot 10^{-\mathrm{dyn}_\mathrm{dB}/10}
\label{Eq: AM excess noise}
\end{eqnarray}
where $E_\mathrm{max}$ is the maximal amplitude to be modulated. The finite dynamics of Alice's AM thus adds a noise proportional to the amplitude $E_\mathrm{max}$. This imperfection is then a limitation to the maximal amplitude of the phase reference pulses in LLO-based CV-QKD designs.\\

\paragraph{Linearity range of the reception detector.} In practice, Bob's detector response is linear with the input number of photons within a finite range. Beyond a threshold, the output of the detector is no longer linear and the security can be broken \cite{SaturationAttackHao}. Thereby, this threshold can be seen as a limitation on the amplitude of the reference pulse used to transmit the phase reference. However, as discussed in Annex.~\ref{Annex:Linearity_Range_Detector}, typical values of this threshold are sufficiently large to allow precise relative phase estimation and, thus, are not a limitation to the reference amplitude.\\

\paragraph{Technical noise.} In order to simulate the experimental imperfections that one can not calibrate within a typical CV-QKD implementations, we introduce a technical excess noise which is typically a fraction of the shot noise ($\xi_\mathrm{tech} = 0.01 N_0$).\\

The necessity of exchanging both weak and intense optical signals in CV-QKD based on a local local oscillator using only one experimental setup is limited by the finite AM dynamics. These effects are not considered in standard TLO designs because only weak quantum signals are modulated and detected but we will show that they are key parameters in order to compare realistic implementations of LLO-based CV-QKD in terms of secret key rate. To our knowledge, the amplitude modulator and linearity range issues have not been taken into account so far in CV-QKD analysis. Based on this refined model, we first analyze practical limitations of the LLO-sequential design and, then, we compare the LLO-sequential implementations with our newly proposed designs.

\section{Towards improved phase reference sharing designs }
\label{Sec: Towards self-coherence}

In this section, we highlight limitations of the LLO-sequential design in terms of tolerable phase noise due to the underlying phase reference sharing scheme. We then propose the novel idea of self-coherence to go beyond that phase noise limit.

\subsection{Limitations of the LLO-sequential design}
\label{Sec: Fundamental limit on LLO-seq}


Since signal and reference pulses follow the same optical path, the estimation process in the LLO-sequential design can be schematically shown using Fig.~\ref{Figure: General phase estimation process} where $\theta_\mathrm{S}^\mathrm{ch}=\theta_\mathrm{R}^\mathrm{ch}$. Thus, we have $V_\mathrm{channel} \approx 0$ and the phase noise stems from two contributions: $ V_\mathrm{est}  =  V_\mathrm{drift} + V_\mathrm{error}$.

One important motivation of the present work is related to the fact that there exists a minimal amount of phase noise $V_\mathrm{est}$ (Eq.~\ref{Eq: Total remaining phase noise}) that can be reached with the LLO-sequential design. The main limitation is due to the fact that signal and reference pulses are emitted with a time delay $1/f$ , leading to a phase noise that cannot be compensated, of variance 
\begin{eqnarray}
V_\mathrm{drift} & = & 2\pi \cdot \frac{\Delta \nu_A + \Delta \nu_B}{f}
\end{eqnarray} 

The phase variance associated to reference pulse phase estimation error, $V_\mathrm{error}$ can be minimized by choosing the amplitude $E_\mathrm{R}$ as large as possible. However the value $E_\mathrm{R}$ that can be chosen in practice is limited by the finite dynamics of her amplitude modulator, and the existence of an associated optical leakage, whose excess noise $\xi_\mathrm{AM}$ is proportional to the amplitude $E_\mathrm{R}$ as discussed in Eq.~\ref{Eq: AM excess noise}. This in practice leads to a compromise regarding the value of $E_\mathrm{R}$, in order to minimize the total excess noise.



The excess noise due to imperfect phase reference sharing reads as (Eq.~\ref{Eq: GMCS Phase noise}):
\begin{eqnarray}
\xi_\mathrm{phase} & = & 2V_A\cdot(1-e^{-V_\mathrm{est}/2})
\end{eqnarray} 
In the regimes of low $V_\mathrm{est}$, it simplifies to 
$\xi_\mathrm{phase} =  V_A \cdot V_\mathrm{est}$. In order to ensure a tolerable value $\xi_\mathrm{phase} \leq 0.1$ (typical value of the null key threshold according to security proofs \cite{SaturationAttackHao,ExpCVQKD-chinois2}),this imposes that $V_\mathrm{drift} \lesssim 0.1/V_A$.


\begin{figure}
\includegraphics[scale=0.7,trim = 0cm 9.5cm 0cm 9.5cm, clip]{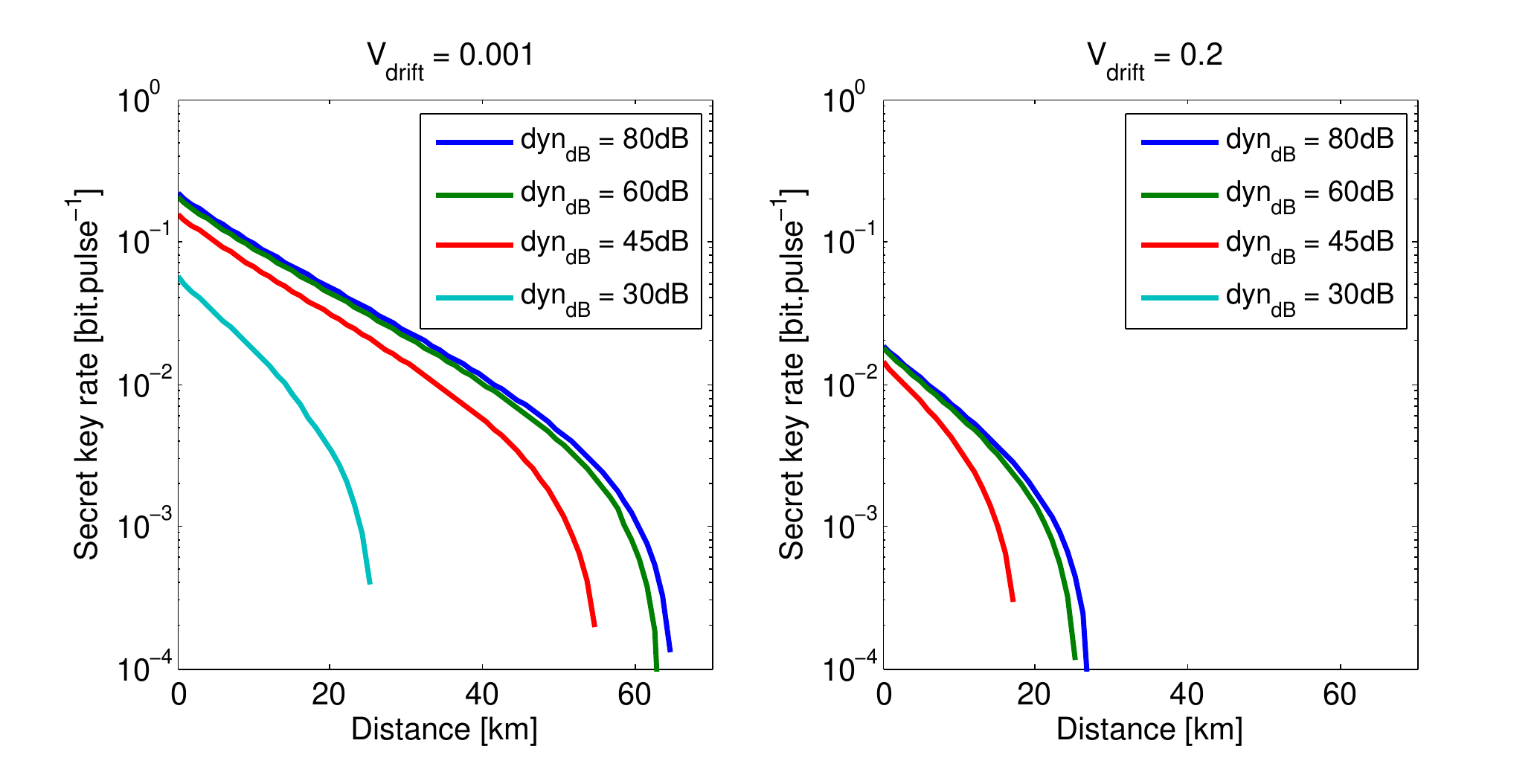} 
\caption{\textit{(color) Secret key rates of the LLO-sequential design for two different values of $V_\mathrm{drift}$ in presence of AM finite dynamics. Simulations are performed in the individual attacks and pessimistic (Eve can control Bob's detection) model \cite{LLOexp2}. The values of $V_A$ and $E_\mathrm{R}$ are chosen to optimize the secret key rate with $\beta = 0.95$, $\eta = 0.7$, $v_{elec} = 0.01$ and $\xi_\mathrm{tech} = 0.01$.}}
\label{Curves:LLO_TolerablePhaseNoise} 
\end{figure}

We can finally express a lower bound on the total excess noise, sum of the excess noise $\xi_\mathrm{phase}$ and the excess noise $\xi_\mathrm{AM}$ in the LLO-sequential design as:
\begin{eqnarray}
\xi_\mathrm{phase} + \xi_\mathrm{AM} & \geq & V_A \cdot \left( V_\mathrm{drift} + \frac{\chi + 1}{E_\mathrm{R}^2} \right) + E_\mathrm{R}^2 \cdot 10^{-\mathrm{dyn}_\mathrm{dB}/10} 
\label{Eq: lower bound}
\end{eqnarray}

We can quantitatively understand from Eq.~\ref{Eq: lower bound} that increasing the amplitude $E_\mathrm{R}$ can reduce the $\xi_\mathrm{phase}$ contribution at the cost of increasing the $\xi_\mathrm{AM}$ contribution. The LLO-sequential design thus requires to the experimental regimes where $V_\mathrm{drift} \ll 1$ and where the amplitude modulator dynamics is large. In other words, the LLO-sequential has to be implemented with performant hardware. 
Current bandwidth limitation of shot-noise limited coherent detectors typically leads to choose $f$ below 100 MHz. This imposes in return strong requirements on the spectral linewidth of the lasers that can be used in order to perform LLO-sequential CV-QKD: the linewidth of the lasers must be at most of $200$ kHz, to ensure an excess noise $\xi_\mathrm{phase} $ lower than $0.1$. As a consequence, only very low phase noise lasers, such as external-cavity lasers (ECL), whose typical spectral linewidth is of a few kHz, are suitable to implement the LLO-sequential design. This is actually illustrated in \cite{LLOexp2}, where the performance analysis is made in the low phase noise regime where $f\gg \Delta \nu_A + \Delta \nu_B$ (ie. in a regime where $V_\mathrm{drift} \approx 0$) and in \cite{LLOexp3} with the experimental choice ultra low noise of ECL lasers of 1.9 kHz linewidth. 

Another issue is actually that finite modulation dynamics has not been taken into account in \cite{LLOexp2}, allowing the authors to choose arbitrary large amplitudes $E_\mathrm{R}$. For instance, they show that, by choosing $E_\mathrm{R}^2 = 500 \ V_A$, a distance of $40$ km is achievable while a more realistic value $E_\mathrm{R}^2 = 20 \cdot V_A$ ($\xi_\mathrm{AM} \approx 10^{-2}$ for $\mathrm{dyn}_\mathrm{dB} = 40$ dB) restricts the protocol to less than $10$ km. This indicated that AM dynamics is an important parameter for analyzing CV-QKD within the LLO framework and the LLO-sequential design requires expensive optical equipments, which is not practical in terms of large-scale deployment for next-generation CV-QKD. 

 In Fig.~\ref{Curves:LLO_TolerablePhaseNoise}, we plot the secret key rates of the LLO-sequential design with finite AM dynamics. We can see that the AM dynamics is an important parameter as it allows to recover the relative phase with good efficiency while ensuring a low excess noise $\xi_\mathrm{AM}$. Below an AM dynamics of $30$ dB, no secret key rate can be produced beyond a distance of around $20$ km, even for a moderate relative phase drift $V_\mathrm{drift} = 10^{-3}$. On the other hand, because of the fundamental limit $V_\mathrm{drift}$ the LLO-sequential design has to be run at a minimal repetition rate to produce secret key, even in large AM dynamics regimes.

In Table.~\ref{Table: Previous work}, we summarize the main characteristics of the two existing implementations of the GMCS protocol proposed so far. Although strong security loopholes have been demonstrated on the TLO implementation, the GMCS protocol has mainly been implemented by directly sending the LO from Alice to Bob. Recent works have however introduced the idea of LLO-based CV-QKD by proposing the experimental LLO-sequential design, hence fixing security weaknesses by generating the LO pulses at Bob side. We have however shown that the LLO-sequential design has strong limitations in terms of implementability in realistic regimes. In the next sections, we investigate how these limitations in term of hardware requirements can be lifted by proposing the idea of self-coherence for phase reference sharing designs. 

\subsection{Self-coherent phase reference sharing schemes}
\label{Sec: Self-coherence}

Performing CV-QKD protocols in the LLO regime can be seen as the issue of conjointly $-$ in the sense of using the same hardware $-$ sharing a phase reference between two remote lasers and performing CV-QKD between the two parties Alice and Bob holding lasers L$_A$ and L$_B$. A first method to perform such task is the LLO-sequential design \cite{LLOexp1,LLOexp2,LLOexp3}. Indeed, the specific modulation of the sequential optical pulses allows one to perform CV-QKD on signal pulses while sharing the phase reference on specific pulses. We have however shown in Sec.~\ref{Sec: Fundamental limit on LLO-seq} a fundamental limitation in terms of tolerable relative phase drift in the LLO-sequential design. As an unspeakable information, the phase reference  has to be encoded over physical carriers, photons in this case. However, by design, the time delay between the emission of quantum signal photons and phase reference photons introduce a decoherence between signal and reference which can prevent any secret key generation in high phase noise regimes.

\begin{table*}
\begin{center}
\begin{tabular}{|c|c|c|c|}
\hline
\textbf{Design} & \textbf{Trusted LO} & \textbf{Tolerable phase noise} & \textbf{Hardware requirements} \\ \hline
Transmitted LO (Fig.~\ref{Scheme: TLO}) &  No & $ \Delta\nu / f \sim 10$ & Stable interferometric set-up \\
\cite{ExpCVQKD-Sequrenet,ExpCVQKD-Qi,ExpCVQKD-chinois,ExpCVQKD-chinois2} & &  &  \\ \hline
LLO-sequential (Fig.~\ref{Scheme: LLO-Sequential}) & Yes & $V_\mathrm{drift} \sim 10^{-1}$ ($60$dB AM) & High AM dynamics \\ 
\cite{LLOexp1,LLOexp2,LLOexp3} & & $V_\mathrm{drift} \sim 10^{-3}$ ($30$dB AM) &  \\ \hline

\end{tabular}
\end{center}
\caption{\textit{Summary of the advantages and drawbacks of all the different CV-QKD designs considered in this work.}}
\label{Table: Previous work}
\end{table*}

We now introduce the novel idea of self-coherence for quantum coherent communication protocols. In order to prevent the phase decorrelation between signal and reference due to sequential emissions, we propose to derive both the signal and reference from the same optical wavefront at emission thus ensuring the physical coherence between signal and phase reference pulses, ensuring that the relative phase drift from Eq.~\ref{Eq: Total remaining phase noise} is $V_\mathrm{drift}=0$. We call \textit{self-coherent} such a design. The relative phase between signal and phase reference is then not affected by the relative phase drift of the lasers and a stable relation between the quantum signal and the LO phases at reception can be provided. As the relative phase estimation does no longer depend on the relative phase drift, self-coherent designs allow Alice and Bob to perform more efficient phase reference sharings. This new method however comes with the challenge of coherently sending $-$ ie. by conserving the stable phase relation $-$ the quantum signal and the phase reference from Alice to Bob. This challenge can be seen as a multiplexing issue. In the remaining of this work, we propose two designs to realize GMCS CV-QKD relying on such self-coherent phase reference sharing designs.

Our first proposal to implement self-coherent CV-QKD is to split a single optical pulse into two pulses used to respectively carry signal and reference information. As output of the same optical pulse, the relative phase between the two pulses at reception only depends on the phases accumulated on their optical paths between emission and reception. This phase reference sharing design relies on the balancing of remote delay line interferometers and we refer to it as the LLO-delayline design. We describe and study its performance in Sec.~\ref{Sec: LLO-delayline}.

A second idea, that we first introduced in \cite{MarieAlleaume2016}, to directly ensure self-coherence between signal and phase reference is to encode both of them within the same optical pulse at emission while recovering both information at reception. In Sec.~\ref{Sec: LLO-displacement}, we propose such a design, the LLO-displacement, in which the phase reference information is encoded over a displacement of the quantum signal modulation. Altough we show that LLO-displacement is restricted to low phase noise, an advantage of this design is that the experimental setup is drastically simplified compare to LLO-delayline, which is a major advantage in the optics of the integration of LLO-based CV-QKD.

\section{Self coherent design based on delay-line interferometer}
\label{Sec: LLO-delayline}

The idea of the LLO-delayline design is to derive consecutive pulse pairs with fixed relative phase, using a balanced delay line interferometer, hence ensuring a self-coherence property. This design does not suffer from the drift limitation of LLO-sequential and can allow Bob to recover the relative phase with better precision.\\

\begin{figure*}
\begin{center}
\includegraphics[scale=0.6,trim = 0cm 4cm 2cm 0.5cm, clip]{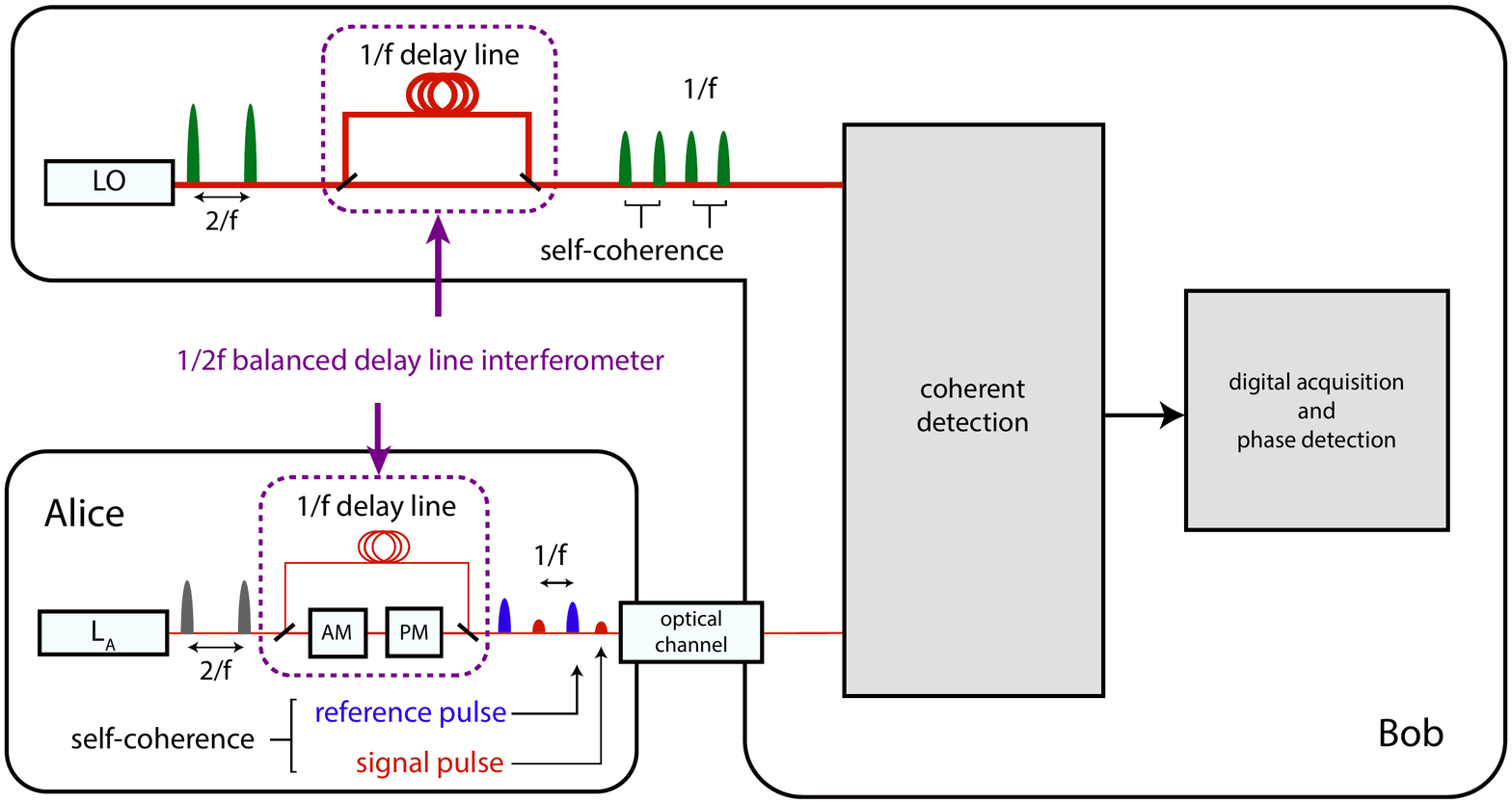}
\caption{\textit{(color) Full experimental scheme of the LLO-delayline design. Alice sends consecutives phase coherent signal/reference pulses pairs to Bob based on a balanced delay line interferometer. On his side, Bob uses his own laser as the LO for coherent detections using the same delay line technique to produce phase coherent LO pulses. Phase estimation and phase correction are digitally performed after measurement acquisition.}}
\label{fig:scheme LLO-delayline}
\end{center}
\end{figure*}

\textbf{The protocol.} The protocol can be decomposed in successive cycles at the repetition rate $f/2$. We note $2\tau = 2/f$ the time interval between two consecutive cycles. Each cycle consists in producing and measuring a self-coherent pair of pulses: one quantum signal pulse and one phase reference pulse. We here describe the protocol for one cycle while Fig.~\ref{fig:scheme LLO-delayline} details the overall design.

At the beginning of a cycle, Alice produces a coherent state $|\alpha _\mathrm{source} \rangle$ which has an optical phase $\varphi_\mathrm{source}^A$. From that single optical pulse, she derives two coherent optical pulses in the following way: she splits the state $|\alpha _\mathrm{source} \rangle$ into two optical pulses, using an unbalanced delayline interferometer:
\begin{itemize}
\item[•] The weak pulse $|\alpha _\mathrm{S} \rangle$ is modulated as the GMCS quantum signal and propagates through an optical path of length $l_A$.
\item[•] The strong pulse $|\alpha _\mathrm{R} \rangle$ is delayed by a time $\tau = 1/f$ on a optical path of length $l_A + \delta l_A$ and is referred to as the reference pulse. 
\end{itemize}

Alice then recombines the two pulses $|\alpha _\mathrm{S} \rangle$ and $|\alpha _\mathrm{R}  \rangle$ resulting in consecutive optical pulses. A major point is that the relative phase between phase reference and quantum signal does no longer depend on the phase drift between Alice and Bob's lasers. An other advantage of this scheme is that the amplitude modulator only modulates the quantum signal which removes the constraints on the AM dynamics. The two optical pulses are then successively sent to Bob through the optical channel, resulting in a repetition rate $f$.

At reception, Bob produces coherent LO pulse pairs using a similar delay line technique used at Alice side. He produces an optical pulse $|\beta _\mathrm{source} \rangle$ with phase $\varphi_\mathrm{source}^B$ and derives two pulses on a 50/50 beamsplitter:
\begin{itemize}
\item[•] The pulse $|\beta _\mathrm{S} \rangle$ that goes through an optical path of length $l_B$.
\item[•] The pulse $|\beta _\mathrm{R} \rangle$ that is delayed and follows an optical path of length $l_B + \delta l_B$.
\end{itemize}
Bob uses the $|\beta _\mathrm{S} \rangle$ and $|\beta _\mathrm{R} \rangle$ pulses as LO pulses to successively measure the received $|\alpha _\mathrm{S} \rangle$ and $|\alpha _\mathrm{R} \rangle$ pulses. This experimental setup can thus be seen as a remote delay-line interferometer split between Alice and Bob sides. 

The reference pulse measurement outcomes allows Bob to calculate an estimation $\hat{\theta}_\mathrm{R}$ of the relative phase $\theta_\mathrm{R}$ at reference measurement and, thus, infer an estimation of the relative phase $\theta_\mathrm{S}$ at signal measurement. Alice can then correct her data to decrease the induced excess noise according to Eq.~\ref{Eq: Alice phase correction}. \\

\textbf{Excess noise evaluation.} In order to study the performance of the LLO-delayline design and calculate the achievable secret key rate, one can note that Alice modulates the quantum signal according to the standard GMCS modulation. Thus, the usual secret key rates formulas of \citep{FossierThesis} can be used. We then have to express the excess noise of the propagating channel in this design, in particular the amplitude modulator noise and the remaining relative phase noise $V_\mathrm{est}$. 

In the LLO-delayline design, the finite dynamics of Alice's amplitude modulator only induces a small contribution to the excess noise $\xi$ in affordable hardware regimes. As it only modulates the quantum signal, the maximal amplitude $E_\mathrm{max}$ of Eq.~\ref{Eq: AM excess noise} does not depend on the reference pulse amplitude. The excess noise $\xi_\mathrm{AM}$ is then independant of the reference amplitude $E_\mathrm{R}$ and the intensity $E_\mathrm{max}^2$ of Eq.~\ref{Eq: AM excess noise} only has to be a few times larger than $V_A$ \cite{FossierThesis}, resulting in a moderate contribution of $\xi_\mathrm{AM}$ to the excess noise $\xi$ ($\xi_\mathrm{AM} \sim 10^{-2}$ for $V_A = 4$, $E_\mathrm{max}^2=10 \ V_A$ and dyn$_\mathrm{dB} = 30$). 

We now quantify the excess noise $\xi_\mathrm{phase}$ by expressing the remaining phase noise $V_\mathrm{est}$ on Bob's estimation of the relative phase. By design, the simultaneous emission on the source pulse of $|\alpha _\mathrm{S} \rangle$ and $|\alpha _\mathrm{R} \rangle$, ie. $t_\mathrm{S}=t_\mathrm{R}$, implies $\theta_\mathrm{drift} = 0$ and, thus, $V_\mathrm{drift}=0$, which corresponds to the self-coherence property. The variance $V_\mathrm{est}$ can then be written as the sum of the phase estimation efficiency $V_\mathrm{error}$ (given in Eq.~\ref{Eq: Single pulse phase estimation}) and the variance $V_\mathrm{channel}$ (Eq.~\ref{Eq: Vchannel}) of the difference between the accumulated phases on the channel: 
\begin{eqnarray}
V_\mathrm{est} & = & V_\mathrm{error} + V_\mathrm{channel}
\label{Eq: Remaining phase noise DLI}
\end{eqnarray}
In this design, signal and phase reference pulses propagate through different optical path. Then, the former term depends on the stability of the delayline interferometer. As introduced in Sec.~\ref{Sec: Noise model}, this corresponds to the variance:
\begin{eqnarray}
V_\mathrm{channel} & = & \mathrm{var}\left( \theta_\mathrm{R}^\mathrm{ch} - \theta_\mathrm{S}^\mathrm{ch} \right)
\end{eqnarray}

where $\theta_\mathrm{S}^\mathrm{ch}$ and $\theta_\mathrm{R}^\mathrm{ch}$ respectively correspond to the phases accumulated by $|\alpha_\mathrm{S}\rangle$ and $|\alpha_\mathrm{R}\rangle$ through their propagation. Therefore, one wants to express the quantity $\theta_\mathrm{channel} =\theta_\mathrm{R}^\mathrm{ch} - \theta_\mathrm{S}^\mathrm{ch}$. Using the definition of the relative phase of Eq.~\ref{Eq: Relative Phase}, we can first write the relative phase as the difference between the phases respectively accumulated by the two interfering LO and signal pulses:
\begin{eqnarray}
\begin{array}{rcl}
\theta_\mathrm{S}^\mathrm{ch} & = & \varphi_{\beta ,\mathrm{S}}^\mathrm{acc} - \varphi_{\alpha, \mathrm{S}}^\mathrm{acc} \\
\theta_\mathrm{R}^\mathrm{ch} & = & \varphi_{\beta , \mathrm{R}}^\mathrm{acc} - \varphi_{\alpha , \mathrm{R}}^\mathrm{acc}
\label{Eq: Accumulated phase}
\end{array}
\end{eqnarray}

where, for instance, $\varphi_{\beta ,\mathrm{S}}^\mathrm{acc}$ stands for the phase accumulated by the LO pulse $|\beta_\mathrm{S} \rangle$ during its propagation. We model the accumulated phase as a linear function $\varphi^\mathrm{acc}(l)$ of the optical path length $l$, then we can derive the following expressions:
\begin{eqnarray}
\begin{array}{rcl}
\varphi_{\alpha , \mathrm{S}}^\mathrm{acc} & = & \varphi^\mathrm{acc}(l_A) \\
\varphi_{\alpha , \mathrm{R}}^\mathrm{acc} & = & \varphi^\mathrm{acc}(l_A) + \varphi^\mathrm{acc}(\delta l_A) \\
\varphi_{\beta , \mathrm{S}}^\mathrm{acc} & = & \varphi^\mathrm{acc}(l_B) \\
\varphi_{\beta , \mathrm{R}}^\mathrm{acc} & = & \varphi^\mathrm{acc}(l_B) + \varphi^\mathrm{acc}(\delta l_B)
\end{array}
\end{eqnarray}

Using the previous equations, we can finally express $\theta_\mathrm{channel} = \theta_\mathrm{R}^\mathrm{ch} - \theta_\mathrm{S}^\mathrm{ch}$  as:
\begin{eqnarray}
\theta_\mathrm{channel} & = & \varphi^\mathrm{acc}(\delta l_B) - \varphi^\mathrm{acc}(\delta l_A)
\label{Eq: Relative phase drift dli}
\end{eqnarray}

As we can see, the relative phase drift $\theta_\mathrm{channel}$ only depends on the difference of the accumulated phases between the delayline optical paths $\delta  l_A$ and $\delta  l_B$. Due to experimental imperfections as thermal fluctuations, we model $\delta l_A$ and $\delta  l_B$ as a stochastic processes over time around the same mean value $\langle \delta l_A \rangle = \langle \delta l _B \rangle = c\tau$ ($c$ being the speed of the light). The phase $\theta_\mathrm{channel}$ then only depends on the fluctuations of the processes $\delta l_A$ and $\delta  l_B$ corresponding to the interferometer balancing efficiency. An important point is that previous experimental demonstrations of CV-QKD \cite{ExpCVQKD-Qi,ExpCVQKD-Sequrenet,ExpCVQKD-chinois2} with transmitted LO, that rely on such delay line interferometers, have proven that phase fluctuations (with frequency typically of order of Hz) associated to interferometer path length fluctuations can be kept low in frequency and amplitude when sampled at CV-QKD repetition rate and do not prevent to perform CV-QKD with repetition rates in the MHz (or above) and we consider that $V_\mathrm{channel}= \mathrm{var}(\theta_\mathrm{channel}) \approx 0$. We can then consider that the variance $V_\mathrm{est}$ from Eq.~\ref{Eq: Remaining phase noise DLI} is dominated by the phase measurement efficiency:
\begin{eqnarray}
V_\mathrm{est} & = & V_\mathrm{error} 
\end{eqnarray}

The LLO-delayline design ensures self-coherence at interference using delayline interferometers and the dependence on the relative phase drift of the lasers is removed. Bob gets self-coherent outcome measurements and is able to estimate the phase drift in the same way as in the LLO-sequential design but with higher efficiency. Finally, this results in the following excess noise:
\begin{eqnarray}
\xi_\mathrm{phase} & = & 2V_A\cdot (1 - e^{-V_\mathrm{error}/2}) 
\end{eqnarray}

\textbf{Performance analysis.} Based on the previous excess noise analysis, we can now study the performance of the LLO-delayline design in terms of secret key rate and compare its performance with the LLO-sequential design. As the quantum signal modulation is the same as in the LLO-sequential, we can equivalently compare the achievable secret key rate or the excess noise contributions.

The LLO-delayline design allows to remove the relative phase drift $V_\mathrm{drift}$ from the excess noise expression. However, the relative phase drift between the two lasers should be stable within the duration of a single optical pulse and imposes that $V_\mathrm{drift} \lesssim 10$ (Sec.~\ref{Sec: TLO}). The remaining phase noise is only limited by the efficiency $V_\mathrm{error}$ of the phase reference estimation. In particular, the LLO-delayline allows to perform CV-QKD stronger phase noise regime than the LLO-sequential design. Furthermore, as the amplitude modulator excess noise $\xi_\mathrm{AM}$ does not depend on the phase reference amplitude, the AM dynamics do no longer restrict the relative phase measurement efficiency $V_\mathrm{error}$. The reference amplitude $E_\mathrm{R}$ can then be chosen as large as possible in the limit of the saturation limit of Bob's detector and of the launched power limit without increasing the excess noise $\xi_\mathrm{AM}$. In practice, these limits allow an very efficient phase measurement.

\begin{figure*}
\begin{center}
\includegraphics[scale=0.75,trim = 0cm 11cm 0cm 11cm, clip]{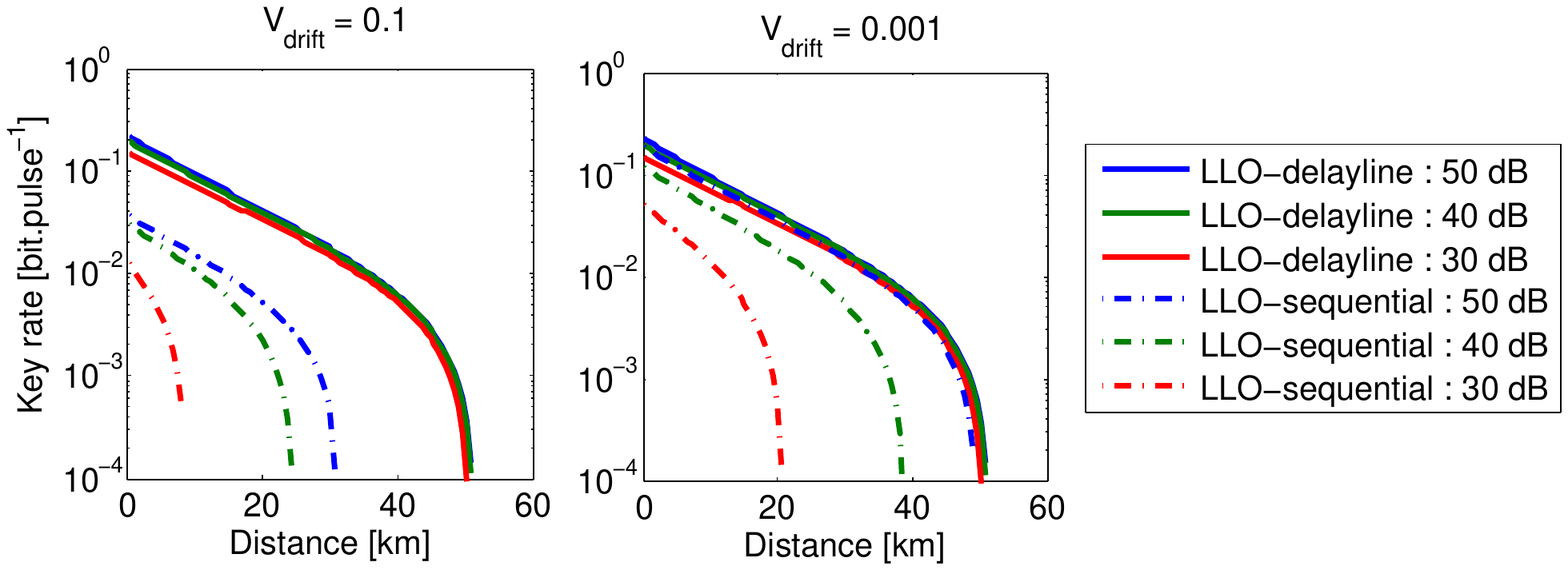}
\caption{\textit{(color) Secret key rate comparaison between the LLO-sequential and the LLO-delayline designs for different AM dynamics and relative phase drift $V_\mathrm{drift}$.}}
\label{Curves: LLO-delayline}
\end{center}
\end{figure*}

In Fig.~\ref{Curves: LLO-delayline}, we plot the expected key rates for both the LLO-sequential and LLO-delayline designs for different relative phase drift and AM dynamics. We can see that the LLO-delayline design is more resilient to both a decrease of the AM dynamics and to an increase of the relative phase drift. Tthe self-coherence between quantum signal and phase reference allows to reach stronger phase noise regime than LLO-sequential with similar optical hardwares. We can see on the left figure of Fig.~\ref{Curves: LLO-delayline} that with a $50$ dB AM, the LLO-delayline design allows secret key generation at $50$ km with $V_\mathrm{drift} = 0.1$ when LLO-sequential can only reach $30$ km. Furthermore, in the LLO-delayline design, the amplitude of the reference pulses can be chosen much larger than in the LLO-sequential design because it does not increse the excess noise $\xi_\mathrm{AM}$. Thus, the relative phase is estimated with better precision reducing the induced excess noise. For instance, the LLO-delayline allows to perform CV-QKD at a distance of $50$ km with affordable $30$ dB amplitude modulators even in the regime of standard DFB lasers (linewidths of order of MHz) which is not possible with the LLO-sequential design. We have thus shown that the LLO-delayline allows to perform LLO-based CV-QKD in the regime of affordable optical hardware regimes, which is an improvement towards LLO-based CV-QKD based on standard optical equipments. 

\section{Self coherent designs based on a modulation displacement}
\label{Sec: LLO-displacement}

In this section, we propose a second design (firstly proposed in \cite{MarieAlleaume2016}), the LLO-displacement design, implementing CV-QKD with a self-coherent phase reference sharing. This design is based on a method for jointly encoding both quantum signal and phase reference information over each optical pulse produced by Alice's laser L$_\mathrm{A}$ at emission.\\ 

\textbf{The protocol.} The main idea is to displace, in the phase space, the modulation sent by Alice with a fixed displacement of amplitude $\Delta$ and phase $\phi_\Delta$. Given her standard GMCS variables $(x_A,p_A)$, Alice produces and sends to Bob the displaced coherent state:
\begin{eqnarray}
|\alpha \rangle _{\overrightarrow{\Delta}} & = & |x_A + \Delta \cos\phi_\Delta, p_A + \Delta \sin\phi_\Delta \rangle
\end{eqnarray}
The amplitude $\Delta$ and the phase $\phi_\Delta$ of the displacement are publicly known so that it carries information on Alice's phase reference. At reception, Bob measures both $\hat{x}$ and $\hat{p}$ quadratures of each received optical pulse using a heterodyne detection and gets measurement outcomes $(x_{B},p_{B})$:
\begin{eqnarray}
\begin{array}{ccc}
x_{B} & = & \sqrt{\frac{G}{2}} \cdot \left[ (x_A + \Delta\cos\theta_\Delta) \cdot \cos\theta  + (p_A + \Delta\sin\theta_\Delta) \cdot \sin\theta  + x_0 + x_c \right]\\
p_{B} & = & \sqrt{\frac{G}{2}} \cdot \left[ - (x_A + \Delta\cos\theta_\Delta) \cdot \sin\theta  + (p_A + \Delta\sin\theta_\Delta) \cdot \cos\theta  + p_0 + p_c \right]
\end{array}
\end{eqnarray}

Using the displacement of Alice's modulation, Bob is able to measure an estimator $\hat{\theta}_\mathrm{S}$ of the relative phase $\theta_\mathrm{S}$ by using his measurement outcomes $(x_B,p_B)$ as detailed in Sec.~\ref{Sec: Noise model}. Furthermore, using the $i$ indexes for successive pulses, Bob can calculate a more precise estimator $\hat{\theta}_\mathrm{filter}^{(i)}$ by averaging each estimator $\hat{\theta}_\mathrm{S}^{(i)}$ with the previous filtered estimator $\hat{\theta}_\mathrm{filter}^{(i-1)}$ using optimized weighted coefficients. Finally, the estimator $\hat{\theta}_\mathrm{filter}^{(i)}$ allows Alice and Bob to correct their data using Eq.~\ref{Eq: Alice phase correction}.\\

\textbf{Security of the protocol.} In order to study the security of the design LLO-displacement and calculate the secret key rate, one can observe that security proofs for the GMCS protocols \cite{LeverrierThesis,GarciaPatronThesis} do not rely on the mean value of Alice's quadrature because it is fully described using the covariance matrices formalism. However, as we will show, the excess noise induced by the phase noise on a displaced modulation is asymmetric. The secret key rates then has to be calculated using a specific method which is detailed in Annex.~\ref{Annex:secret key rate formulas}.

We now quantify the remaining phase noise $V_\mathrm{est}$ of Eq.~\ref{Eq: Total remaining phase noise}. As both quantum signal and phase reference are encoded and transmitted within the same optical pulse, we can directly write $V_\mathrm{drift}=0$ and $V_\mathrm{channel} = 0$. Finally, the only term contributing to the remaining phase noise $V_\mathrm{est}$ is the variance $V_\mathrm{error}$. Due to the the particular modulation scheme however, the variance of the estimates $\hat{\theta}_\mathrm{S}^{(i)}$ is not expressed as Eq.~\ref{Eq: Vchannel}. In this case, Alice's modulation can be seen as a noise in the phase estimation process, resulting in:
\begin{eqnarray}
V_\mathrm{error} & = & \frac{V_A + \chi + 1}{\Delta^2}
\end{eqnarray}
Furthermore, the filtering technique $-$ hence based on all previous relative phase estimates $-$ allows to use correlations of the phase drift over time to recover the relative phase with better precision than $V_\mathrm{error}$. In the asymptotic regime (when $i$ is large), we can show that the successive variances $V_\mathrm{filter}^{(i)}$ tend to an asymptotic limit and, finally, one can write: 
\begin{eqnarray}
V_\mathrm{est} & = & \sqrt{V_\mathrm{error}} \cdot \sqrt{V_\mathrm{drift}}
\label{Eq: optimal filter size 1 approx}
\end{eqnarray}
where $V_\mathrm{drift}$ is the relative phase drift between lasers L$_A$ and L$_B$ between two consecutive pulses, ie. it is expressed as Eq.~\ref{Eq: Relative phase drift LLO-sequential}.

We can now express the excess noise $\xi_\mathrm{phase}$ due to phase noise in the LLO-displacement regime (using Annex.~\ref{Annex: Phase excess noise}). It can be simplified when $V_\mathrm{est} \ll 1$ and reads as:
\begin{eqnarray}
\begin{array}{ccl}
\xi_\mathrm{phase}^{(x)} & = & (V_A + \Delta^2 \cdot \sin^2 \phi_\Delta ) \cdot V_\mathrm{est} \\
\xi_\mathrm{phase}^{(p)} & = & (V_A + \Delta^2 \cdot \cos^2 \phi_\Delta ) \cdot V_\mathrm{est}
\label{Eq: Phase Excess Noise Displacement}
\end{array}
\end{eqnarray}

\begin{figure*}
\begin{center}
\includegraphics[scale=0.75,trim = 0cm 11cm 0cm 11cm, clip]{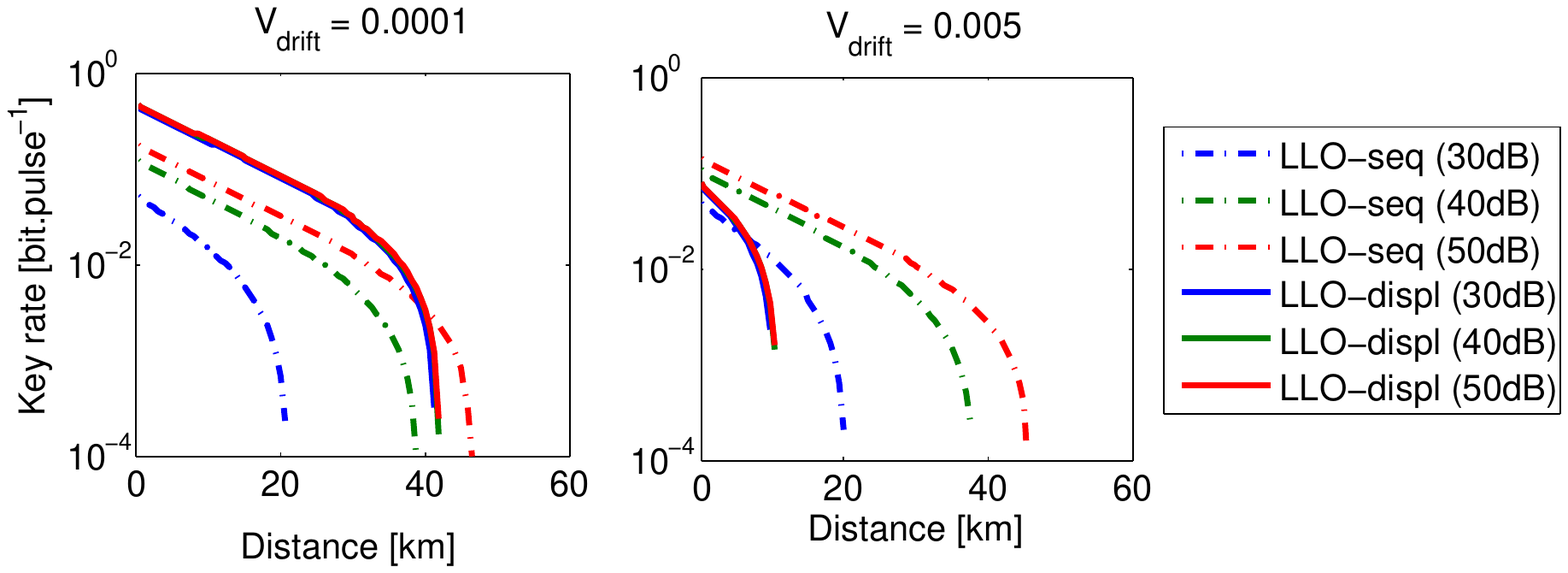}
\caption{\textit{(color) Secret key rate comparaison between the LLO-sequential and the LLO-displacement designs for two relative phase drift $V_\mathrm{drift}$ and for different AM dynamics.}}
\label{Curves: LLO-displacement}
\end{center}
\end{figure*}

where the expression of $V_\mathrm{est}$ is defined using Eq.~\ref{Eq: optimal filter size 1 approx}. A crucial point in the noise analysis of the LLO-displacement design is that the displacement of Alice's modulation creates an asymmetry on the excess noise on each quadrature. For instance, if $\phi_\Delta = 0$ (displacement according to the $\hat{x}$ quadrature), the displacement induces an increasing of the excess noise on the $\hat{p}$ quadrature.

Furthermore, in this design, the maximal amplitude $E_\mathrm{max}$ of the AM excess noise (Eq.~\ref{Eq: AM excess noise}) can be approximate as $\sqrt{V_A + \Delta^2}$, resulting in: 
\begin{eqnarray}
\xi_{AM} & = & (V_A + \Delta^2) \cdot 10^{-\mathrm{dyn}_\mathrm{dB}}
\end{eqnarray}

However, one can note that the excess noise $\xi_\mathrm{phase}$ is a more restricting limitation to the displacement amplitude than the excess noise $\xi_{AM}$ for realistic parameters and, in practice, the AM dynamics do not restrict the amplitude displacement.\\

\textbf{Performance analysis.} In \cite{MarieAlleaume2016}, we only considered the $\xi_\mathrm{phase}^{(x)}$ contribution in the case of $\phi_\Delta =0$, resulting in a too optimistic key rate. Although the displacement decreases the variance $V_\mathrm{error}$ and, thus, the remaining phase noise $V_\mathrm{est}$, it also increases its impact on the excess noise according to Eq.~\ref{Eq: Phase Excess Noise Displacement}. Unfortunately, this result is a strong limitation to the achievable $\Delta$ and, thus, to the tolerable phase noise in the LLO-displacement design. 

We here consider the case where $\phi_\Delta = 0$ (other cases can be treated in a similar way by observing that the sum $\xi_\mathrm{phase}^{(x)}+\xi_\mathrm{phase}^{(p)}$ does not depend on $\phi_\Delta$). From Eq.~\ref{Eq: Phase Excess Noise Displacement}, one can note that the phase excess noise induced on the $\hat{p}$ quadrature is proportional to the displacement mean photon number $\Delta^2$. This sets a strong constraints on the achievable value of $\Delta$ and, thereby, on the achievable value of $V_\mathrm{error}$. There is a trade-off in terms of the $\Delta$ value between the remaining phase noise $V_\mathrm{est}$ $-$ the displacement $\Delta$ decreases $V_\mathrm{error}$ $-$ and the  excess noise $\xi_\mathrm{phase}^{(p)}$. An optimal value for the displacement can be found and is calculated in our simulations. However, the optimal value of the $\Delta$ only allows secret key generation for low values of $V_\mathrm{drift}$. This means that, solely based on the single estimates $\hat{\theta}_\mathrm{S}$, the LLO-displacement design does not allow a low enough excess noise $\xi_\mathrm{phase}$ and, as such, requires strong correlations, ie. low values of $V_\mathrm{drift}$, between consecutive relative phases to recover phase information based on filtering techniques.

In Fig.~\ref{Curves: LLO-displacement}, we plot the expected key rates for both the LLO-sequential and the LLO-displacement designs. The displacement value is optimized to maximize the overall secret key rate. In low phase noise regimes, we can see that the LLO-displacement is better in terms of secret key rate generation. This is due to the fact that the LLO-displacement relies on the whole repetition rate to generate secret key and the filtering technique combined to the displacement allows a more efficient relative phase recovery. For higher phase noise regimes however, the displacement value required to estimate the relative phase is limited by the excess noise $\xi_\mathrm{phase}^{(p)}$ and, finally, the relative phase estimation can not be performed with a good efficiency.\\

We have shown that the coherence coming from the simultaneous encoding of both the quantum signal and phase reference in this design comes with new challenges. In particular, the displacement of the modulation increases the excess noise induced by the relative phase noise and creates an asymmetry on the excess noise on each quadratures. An interesting issue is then to optimize the LLO-displacement design in order to increase its performance, especially in terms of phase noise resilience. This study is however kept for future works. We however emphasize that LLO-displacement design relies on an extremely convenient experimental scheme as the phase reference encoding is performed simultaneously with the quantum signal modulation. To our knowledge, the simultaneous quantum signal and phase reference transmission introduced in the LLO-displacement design is a new primitive which has not been studied so far in quantum communication regimes and can be applied to different signal modulations as BPSK or higher order modulations. Since both quantum signal and phase reference information are sent over the same optical pulse, the key rate obtained with the LLO-displacement design is moreover not lowered by time multiplexing, as it is the case with LLO-sequential. Thereby, unlike all other proposals for locally generated local oscillator based CV-QKD designs, it allows to use the whole repetition rate for secret key generation.

\section{Conclusion and perspectives}
\label{Sec: Conclusion}

In order to lift security loophole issues, the local oscillator should not be directly sent through the optical channel in CV-QKD experiments and LLO-based CV-QKD protocols have been introduced. We have however shown that the only design proposed to date, the LLO-sequential design \cite{LLOexp1,LLOexp2,LLOexp3}, requires to use ultra low noise lasers and high dynamics modulators. This strong requirements in terms of hardware performance are a limitation to the ability to deploy LLO-based CV-QKD over large-scale optical networks. In this work, we have addressed the issue of performing CV-QKD with a local local oscillator, using affordable hardware.

\begin{table*}
\begin{center}
\begin{tabular}{|c|c|c|c|}
\hline
\textbf{Design} & \textbf{Trusted LO} & \textbf{Tolerable phase noise} & \textbf{Hardware requirements} \\ \hline
Transmitted LO  &  & & \\
\cite{ExpCVQKD-Sequrenet,ExpCVQKD-Qi,ExpCVQKD-chinois,ExpCVQKD-chinois2} & No & $\Delta \nu/f \sim 10$ & Stable interferometric set-up \\
(Section ~\ref{Sec: TLO}) & & & \\ \hline
LLO-sequential  & & &  \\ 
\cite{LLOexp1,LLOexp2,LLOexp3} & Yes & $V_\mathrm{drift} \sim 10^{-2}$ & High AM dynamics \\
(Section ~\ref{Sec: LLO-seq}) & & & \\  \hline \hline

LLO-delayline  & & & \\

(Section ~\ref{Sec: LLO-delayline}) & Yes & $V_\mathrm{drift} \sim 10$ & Stable interferometric set-up \\ 
& & & \\ \hline
LLO-displacement &  &  &\\ 
(Section ~\ref{Sec: LLO-displacement}) & Yes & $V_\mathrm{drift} \sim 10^{-4}$ & $\emptyset$ \\ 
& & & \\ \hline

\end{tabular}
\end{center}
\caption{\textit{Summary of all the CV-QKD designs discussed in this work. We compare them in terms of tolerable phase noise and on their experimental limitations.}}
\label{table}
\end{table*}


 The main challenge of LLO CV-QKD is that the phase drift between emitter laser and local oscillator laser, placed at Bob, induces a phase noise on the quantum communication, that has to be efficiently corrected. In Table~\ref{table}, we summarize the performance and requirements of existing as well as newly proposed designs for CV-QKD. The LLO-sequential design is intrinsically limited to low phase noise regimes. This puts important constraints on the type of lasers that can be used both as emitter and LO. An other limitation of the LLO-sequential is the efficiency of the relative phase estimation process, which is limited in practice by Alice's amplitude modulator dynamics.

Our results imply that next generation CV-QKD, implemented with a local LO, is possible even with low cost DFB lasers and standard amplitude modulators. Such features are made possible by the newly introduced self-coherent phase reference sharing design, the LLO-delayline design, and are essential to progress towards photonic integration and wide deployment of CV-QKD.  The LLO-delayline design relies on the self-coherence between the quantum signal and phase reference and on the interferometric stability of two delay lines at short time scale. The finite dynamics of the amplitude modulator does no longer restrict the reference pulse amplitude and, by consequence, the relative phase estimation process. These characteristics allow the LLO-delayline design to be more resilient to phase noise than the previously proposed LLO-sequential design. In Fig.~\ref{Curves: LLO-delayline}, we can see that the LLO-delayline design is able to reach a distance of $50$ km in a regime of high phase noise,$V_\mathrm{drift}=0.1$, while the reachable distance with the LLO-sequential design is below $25$ km even with large AM dynamics. Furthermore, we emphasize that remote delay-line interferometric stability has already been demonstrated in practice on several CV-QKD implementations \cite{ExpCVQKD-Sequrenet,ExpCVQKD-chinois2} paving the way to the demonstration of LLO CV-QKD with cheap hardware, using the LLO-delayline design.

As another contribution, we have investigated a scheme, LLO-displacement, allowing to simultaneously transmit the quantum signal and the phase reference information on the same optical pulse. We have however observed that the implementation of the LLO-displacement design with Gaussian modulated coherente states (GMCS CV-QKD protocol) leads to an overall excess noise that increases with displacement, which restricts its use to low phase noise regimes. Simultaneous transmission of quantum and phase reference information had however not been studied so far and our results can be of interest in view of performing a joint optimization of classical and quantum coherent communication systems, operating with the same hardware. The optimization of such protocols is then an interesting open question and is kept for future works.

\textbf{Aknowledgements.} The authors thank Bing Qi for noticing an important limitation of the LLO-displacement design, in terms of additional excess noise, as well as for the fruitful exchanges and discussions on the LLO-displacement design.

\bibliographystyle{unsrt}
\bibliography{bibliography}

\section{Annex}

\subsection{Excess noise due to phase noise}
\label{Annex: Phase excess noise}

Alice sends the coherent state $|\alpha \rangle = |x_A + x_0, p_A + p_0 \rangle$, where, in the general case, we suppose that $x_A \sim \mathcal{N}(0,V_x)$ and $p_A \sim \mathcal{N}(0,V_p)$ while Bob uses a heterodyne detection at reception. In order to estimate the phase noise induced excess noise, we consider in this analysis that the relative phase noise is the only noise source. Bob then gets the following measurement outcomes $({x}_m,{p}_m)$:
\begin{eqnarray}
\left( \begin{array}{c} {x}_m \\ {p}_m\end{array} \right) & = & \sqrt{\frac{G}{2}} \cdot \left[ \begin{pmatrix} \cos{\theta}  & \sin{\theta} \\ -\sin{\theta}  & \cos{\theta}  \end{pmatrix} \cdot \left( \begin{array}{c} x_A + x_0  \\ p_A + p_0 \end{array} \right) \right]
\end{eqnarray}

We suppose that Bob gets an estimator $\hat{\theta} \sim \mathcal{N}(\theta, V_{\varphi})$. He sends his estimator to Alice which corrects her data and, in the reverse reconciliation scheme, Alice then estimates Bob's measurements as:
\begin{eqnarray}
\begin{pmatrix} \tilde{x_A} \\ \tilde{p_A} \\ \end{pmatrix} & = & \sqrt{\frac{G}{2}} \cdot \begin{pmatrix} \cos \hat{\theta} & \sin \hat{\theta}\\ 
-\sin \hat{\theta} & \cos \hat{\theta}\\ \end{pmatrix} \cdot \begin{pmatrix} x_A + x_0  \\ p_A + p_0 \\ \end{pmatrix}
\label{quad estimator v1}
\end{eqnarray}

We can then express the excess noise on each quadrature as:
\begin{eqnarray}
\begin{array}{rcl}
\xi_x & = & \mathrm{var}(x_m-\tilde{x}_A) \\
\xi_p & = & \mathrm{var}(p_m-\tilde{p}_A)
\end{array}
\end{eqnarray}

These two quantities depends on the remaining relative phase $\varphi = \theta - \hat{\theta}$. Assuming that the variable $\varphi$ is a Gaussian variable such that $\varphi \sim \mathcal{N}(0,V_{\varphi})$, the above expressions can be calculated from the characteristic function of the Gaussian function and, after calculations, we obtain the following expressions:
\begin{eqnarray}
\begin{array}{ccc}
\xi_\mathrm{phase}^{(x)} & = & V_x \cdot (1+e^{-V_\varphi}-2e^{-V_\varphi /2}) + (V_x + x_0^2)\cdot(\frac{1}{2}+\frac{1}{2}e^{-2V_\varphi} - e^{-V_\varphi}) + (V_p + p_0^2)\cdot(\frac{1}{2} - \frac{1}{2} e^{-2V_\varphi})\\ 
\xi_\mathrm{phase}^{(p)} & = & V_p \cdot (1+e^{-V_\varphi}-2e^{-V_\varphi /2}) + (V_p + p_0^2)\cdot(\frac{1}{2}+\frac{1}{2}e^{-2V_\varphi} - e^{-V_\varphi}) + (V_x + x_0^2)\cdot(\frac{1}{2} - \frac{1}{2} e^{-2V_\varphi})
\end{array}
\end{eqnarray}

\subsection{Linearity range of the coherent detector}
\label{Annex:Linearity_Range_Detector}

\begin{figure*}
\includegraphics[scale=0.6,trim = 6cm 9cm 2.5cm 5.2cm, clip]{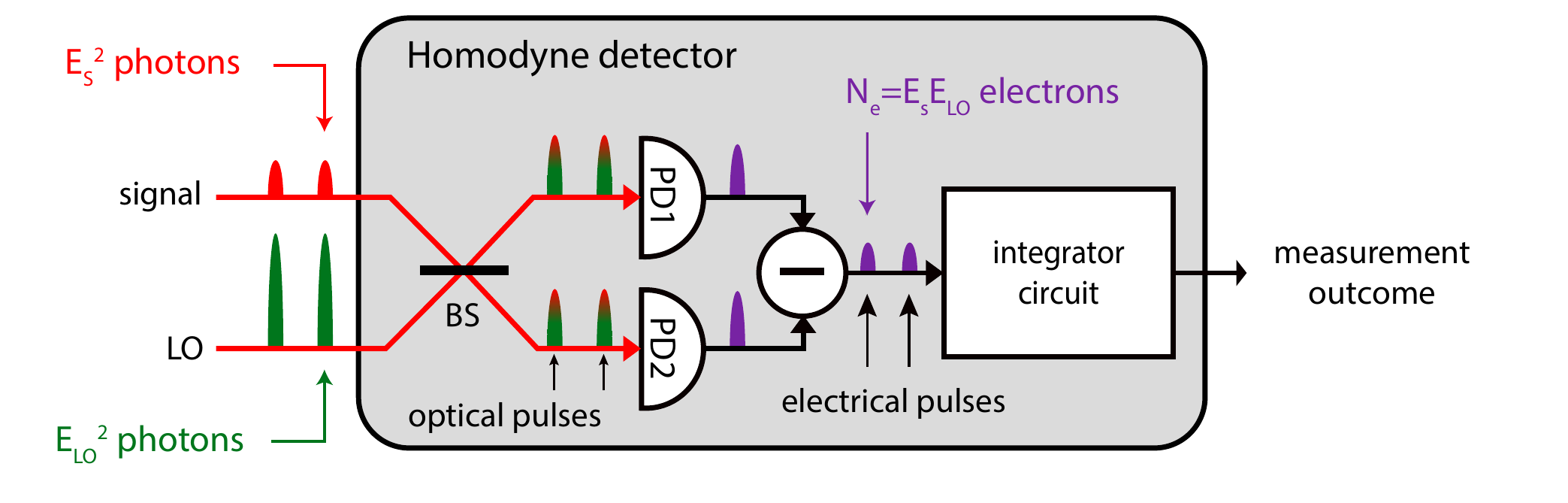}
\caption{\textit{(color) Scheme of a typical homodyne detector. Signal and local oscillator pulses interfer on a 50/50 beamsplitter (BS). Both resulting fields are detected on two photodiodes (PD1 and PD2) which convert photons into electrons. The electrical pulses (purple pulses) produced by the photodiodes are then substracted (-) and the resulting quadrature electrical pulse intensity is measured using a integrator circuit. The outcome of the integrator circuit is proportionnal to $N_\mathrm{e} = E_\mathrm{S} \cdot E_\mathrm{LO}$ up to an intensity threshold $N_\mathrm{sat}$.}}
\label{fig:scheme HD}
\end{figure*}

We consider that Bob relies on a single coherent detector which addresses the issue of the linearity range of the detector when considering both quantum signal and phase reference transmission. A typical homodyne detector is presented in Fig.~\ref{fig:scheme HD}. The response of the integrator circuit is proportional to the incoming number of electrons over a finite range. Beyond a certain threshold, the response of the integrator circuit is no longer linear and the security can be broken by specific attacks \cite{SaturationAttackHao}. We define this threshold as the maximal number of electrons $N_\mathrm{sat}$ per electrical pulses that can be detected in a linear regime. The number of electrons in each electrical pulse is $N_\mathrm{e} = 0.5 \cdot G \cdot E_\mathrm{S} \cdot E_\mathrm{LO}$ (see Fig.~\ref{fig:scheme HD}) where $E_\mathrm{S}$ and $E_\mathrm{LO}$ are the amplitudes of the signal and the LO so that the saturation hypothesis imposes:
\begin{eqnarray}
\frac{G}{2} \cdot  E_\mathrm{S} \cdot E_\mathrm{LO} \leq N_\mathrm{sat}
\label{saturation}
\end{eqnarray}
A example of saturation threshold $N_\mathrm{sat} = 10^6$ has been experimentally evaluated in \cite{SaturationAttackHao}. For quantum signal of intensity of order of the shot noise, this threshold is not important and has not been considered so far in CV-QKD analysis. In LLO-based CV-QKD however, the relatively large amplitude of phase reference pulses imposes to consider the saturation threshold as a limit on the reference pulses amplitude.

\begin{figure*}
\includegraphics[scale=0.6,trim = 2.5cm 10cm 2.5cm 10cm, clip]{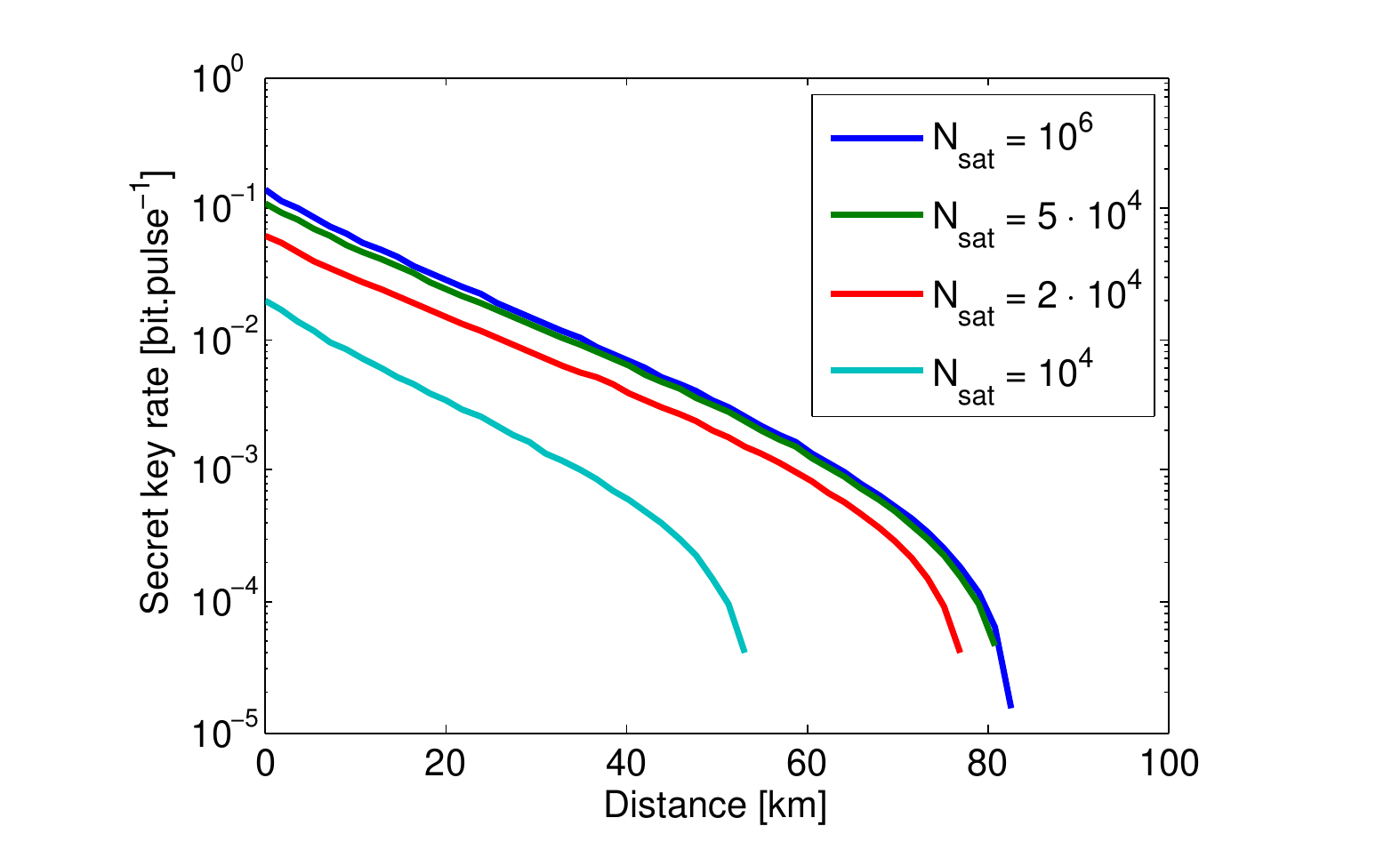}
\caption{\textit{(color) Expected secret key rates for the LLO-sequential design for different value of linearity threshold $N_\mathrm{sat}$.}}
\label{fig:linearity}
\end{figure*}

Eq.~\ref{saturation} implies a trade-off, in the LLO-sequential design, between the signal amplitude $-$ in particular the reference amplitude $E_\mathrm{R}$ $-$ and the local oscillator amplitude $-$ used to decrease the electronic to shot noise ratio. If we want to maximize these two quantities, one has to saturate Eq.~\ref{saturation} by choosing:
\begin{eqnarray}
E_\mathrm{LO} & = & \frac{2}{G} \cdot \frac{N_\mathrm{sat}}{E_\mathrm{R}}
\end{eqnarray}
The electronic to shot noise ratio of Eq.~\ref{Eq: electronic noise} is then written as:
\begin{eqnarray}
\xi_\mathrm{elec} & = & \frac{G\cdot 10^6}{2} \cdot \frac{E_\mathrm{R}^2}{N_\mathrm{sat}^2}
\end{eqnarray}
where $v_\mathrm{elec}=0.01$ and $E_\mathrm{LO,cal}^2 = 10^8$. In Fig.~\ref{fig:linearity}, we plot the expected key rate of the LLO-sequential design for different values of the threshold $N_\mathrm{sat}$. As we can see, only low values of $N_\mathrm{sat}$ (two order of magnitude lower than the experimental value of \cite{SaturationAttackHao}) are limitations to this design. As a typical value of $N_\mathrm{sat}=10^6$ photons is sufficiently large to allow a precise relative phase sharing, the saturation threshold will not be a limitation to the reference amplitude. 

\subsection{Secret key rate formulas for CV-QKD}
\label{Annex:secret key rate formulas}

In this work, we focus on the Gaussian-modulated coherent state (GMCS) protocol. In this protocol, Alice encodes zero-mean gaussian classical variables $x_A$ and $p_A$ on both the $\hat{x}$ and $\hat{p}$ quadratures of coherent states \cite{GaussianQuantumInformation} before sending them to Bob through an insecure optical channel controlled by an eavesdropper Eve. In the Gaussian model, the channel between Alice and Bob is fully characterized by the intensity transmission $G$ and the excess noise $\xi_x$ and $\xi_p$ (\textit{a priori} different on each quadratures. We also assume that Bob uses a heterodyne detection at reception.

\textbf{Symmetric channel.} We first detail the secret key rate formulas used for symmetric excess noise $\xi = \xi_x = \xi_p$. This formulas are used for the LLO-sequential and the LLO-delayline designs. We consider that Eve controls the whole excess noise and we also consider individual attacks. The secret key rates is written as \cite{FossierThesis}:
\begin{eqnarray}
k & = & \beta \cdot I_{AB} - I_{BE}
\label{Eq: Secret key rate ANNEX}
\end{eqnarray}

where:
\begin{eqnarray}
I_{AB} & = & \frac{1}{2} \cdot \log_2 \left( \frac{V+\chi}{1+\chi} \right)
\end{eqnarray}
\begin{eqnarray}
I_{BE} & = & \frac{1}{2} \cdot \log_2 \left( \frac{G\cdot(V+\chi)\cdot(V+\chi_E)}{(\chi_E+1)\cdot(V+1)} \right)
\end{eqnarray}

with:
\begin{eqnarray}
V & = & V_A +1 \\
\chi & = & \frac{2-G}{G} + \xi \\
\chi_E & = & \frac{G\cdot(2-\xi)^2}{(\sqrt{2-2G+G\xi}+\sqrt{\xi})^2}+1
\end{eqnarray}

\textbf{Asymmetric channel.} We have shown in Sec.~\ref{Sec: LLO-displacement} that the excess noise induced by the phase noise in the case of the displaced modulation is asymmetric in the two quadratures. We then need to derive specific secret key rate expressions. From \cite{GarciaPatronThesis,FossierThesis}, we can write:
\begin{eqnarray}
k & = & k_x + k_p
\end{eqnarray} 
where $k_x$ and $k_p$ represent the respective key rates on the logical channel corresponding to each quadrature. Each of these two key rates can then be obtained using the secret key formulas from Eq.~\ref{Eq: Secret key rate ANNEX}, using the corresponding expression  $\xi_x$ and $\xi_p$ from Eq.~\ref{Eq: Phase Excess Noise Displacement}.

\end{document}